\documentclass[smallextended,natbib]{svjour3r}
\smartqed  
\usepackage{graphicx}
\usepackage{array}
\def\thline{\noalign{\hrule height 0.7pt}}
\newtheorem{lm}{Lemma}
\newtheorem{rem}{Remark}
\newtheorem{thm}{Theorem}
\newtheorem{prop}{Proposition}
\newcommand{\1}{\mbox{1}\hspace{-0.25em}\mbox{l}}

\newcommand{\R}{{\mathbb{R}}}
\newcommand{\Z}{{\mathbb{Z}}}
\renewcommand{\S}{{\mathbb{S}}}
\newcommand{\Unif}{{\mathrm{Unif}}}
\newcommand{\Vol}{{\mathrm{Vol}}}

\newcommand{\Cov}{\mathrm{Cov}}
\newcommand{\conv}{\mathrm{conv}}

\usepackage{amsmath,amssymb,amsfonts}

\begin{document}

\title{
Approximate tail probabilities of the maximum of a chi-square field on multi-dimensional lattice points and their applications to detection of loci interactions
}

\author{
Satoshi Kuriki, Yoshiaki Harushima, Hironori Fujisawa, and Nori Kurata
}

\institute{
S.\ Kuriki \at
Institute of Statistical Mathematics,
10-3 Midoricho, Tachikawa, Tokyo 190-8562, Japan.
\email{\texttt{kuriki@ism.ac.jp}}
\and
Y.\ Harushima \at
National Institute of Genetics,
Yata 1111, Mishima, Shizuoka 411-8540, Japan.
\email{\texttt{yharushi@lab.nig.ac.jp}}
\and
H.\ Fujisawa \at
Institute of Statistical Mathematics,
10-3 Midoricho, Tachikawa, Tokyo 190-8562, Japan.
\email{\texttt{fujisawa@ism.ac.jp}}
\and
N.\ Kurata \at
National Institute of Genetics,
Yata 1111, Mishima, Shizuoka 411-8540, Japan.
\email{\texttt{nkurata@lab.nig.ac.jp}}
}

\date{}

\maketitle

\begin{abstract}

Define a chi-square random field on a multi-dimensional lattice points
index set with a direct-product covariance structure,
and consider the distribution of the maximum of this random field.
We provide two approximate formulas for the upper tail probability
of the distribution based on nonlinear renewal theory and
an integral-geometric approach called the volume-of-tube method.
This study is motivated by the detection problem of the interactive
loci pairs 
which play an important role in forming biological species.
The joint distribution of scan statistics for detecting the pairs
is regarded as the chi-square random field above,
and hence the multiplicity-adjusted $p$-value can be calculated
by using the proposed approximate formulas.
By using these formulas,
we examine the data of \citet*{Mizuta-etal10}
who reported a new interactive loci pair of rice inter-subspecies.
\end{abstract}


\section{Introduction}
\label{sec:intro}

\subsection{Tests of multiplicity in detecting loci interactions}

In genomic data analyses, genome scans for detecting loci that have some
particular and interesting functions are often undertaken. 
These procedures are regarded as repeated statistical testings,
and hence they are formalized as multiple testing procedures.
In such multiple testings,
one crucial point is how to adjust the multiplicity of tests.  This is
because the method of adjustment seriously affects the interpretation
of the data analysis.

The detection of the interactive loci pairs
assumed to exist in the Bateson-Dobzhansky-Muller (BDM) model,
which motivates our study,
is such a genome scan problem.
In biological concept,
``species'' are defined as ``groups of interbreeding natural
populations which are reproductively isolated from other such groups''
 (\citet{Mayr42}).
The genetic mechanism for separating species is called reproductive isolation,
which is observed as hybrid sterility or hybrid inviability between particular groups.
The BDM model is a model
for explaining evolution of genetic incompatibility genes.
More precisely, the BDM model assumes that there exist pairs of loci such that
when the loci have particular genotypes,
sterility or inviability occurs and hence a descendant is not produced
 (\citet{Dobzhansky51}, \citet{Coyne-Orr04}).
In this paper, we refer to the interactive loci pair as the BDM pair.

The importance of studying such interactive pair loci is widely acknowledged.
However, few studies have succeeded in identifying
 such pairs and in revealing the mechanism behind them.
For the detection of BDM pairs, choosing two groups used for crossing
is crucial but difficult.
If parents are genetically separate, then descendants cannot be produced.
Conversely,
if parents are too close, then sterility or inviability cannot be observed.
The detection of a BDM pair of 
{\it Arabidopsis\/} intra-species by \citet{Bikard-etal09}, and 
the detection of a BDM pair of rice inter-subspecies by \citet{Mizuta-etal10}
are exceptionally successful studies.

\begin{figure}[h]
\begin{center}
\scalebox{0.95}{\includegraphics{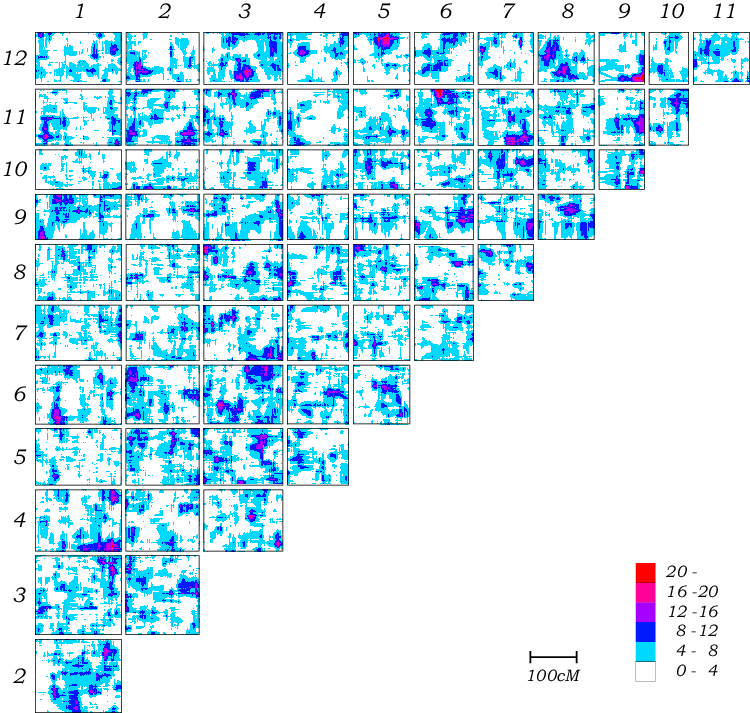}} 
\end{center}
\caption{Contour plot of chi-square statistics}
\label{fig:contour-f2}
\end{figure}

The original purpose of this paper is to give an answer to a statistical
problem that \citet{Mizuta-etal10} have faced in the course of their studies.
Figure \ref{fig:contour-f2} is the contour plot depicting scan statistics
for detecting BDM pairs in a 2nd filial generation ($\mathrm{F}_2$) population
from two rice subspecies used by \citet{Mizuta-etal10}.
The horizontal and vertical axes represent loci positions in 12 chromosomes
of rice.
Each scan statistic is a chi-square statistic with 4 degrees of freedom,
and the number of statistics is around 500,000.
Because of the large number of tests,
some adjustment for the multiplicity of tests is necessary.
The Bonferroni adjustments are frequently used in multiple testing.
However, in our case where the statistics are highly correlated with each other,
the Bonferroni adjustment that is calculated without information of correlation would lead to
very conservative results.

The multiplicity-adjusted $p$-value for correlated scan statistics is defined
from the distribution of their maximum.
For calculating this distribution, we require
knowledge of the correlation structure or joint distribution.
This structure can be determined from experimental design
in the case of crossing experiments such as the detection problem of BDM pairs.
In particular, when the number of statistics is large
and when the correlation structure is systematic,
we can consider a large number of scan statistics as a random field
and can obtain the distribution of the maximum.
The distribution of the maximum of a random field (process)
has been extensively studied.
In this paper, the approaches we use are nonlinear renewal theory and
the volume-of-tube method (tube method).
The nonlinear renewal theory we use was developed by
\citet{Woodroofe82} and \citet{Siegmund85,Siegmund88}. 
In this method, a random field is locally treated as a random walk, and
the distribution of its maximum is obtained by using sequential analysis.
The volume-of-tube method is an integral-geometric approach for approximating
the distribution of the maximum of a Gaussian random field
through evaluating the volume of the index set
 (%
\citet{Sun93},
\citet{Kuriki-Takemura01,Kuriki-Takemura09}%
).
Mathematically, this is equivalent to applying the Euler characteristic heuristic
to a Gaussian field
 (\citet{Takemura-Kuriki02}, \citet{Adler-Taylor07}%
).

This paper is organized as follows.
In Section \ref{subsec:scan},
we explain the scan statistics for detecting BDM pairs.
Under the null hypothesis that a BDM pair does not exist, we see that
the joint distribution of the scan statistics is regarded asymptotically
as a chi-square random field with a direct-product covariance structure
restricted on a lattice point index set.
We also discuss other statistical problems that have the same stochastic
structure as the detection of BDM pairs in Section \ref{subsec:other}.
In Section \ref{sec:maximum}, we formalize this chi-square random field
in a general setting, and provide approximate formulas for
its maximum distribution
by using nonlinear renewal theory and the volume-of-tube method.
Renewal theory assumes that the lattice points are equally spaced.
This assumption may be unreasonable,
because it implies that marker spacings are uniform.
Hence, we use numerical comparisons to examine the difference between the 
randomly spaced case and the equally spaced case.
The volume-of-tube method yields asymptotically conservative bounds
by embedding the random field defined on a discrete set
 (i.e., unequally spaced lattice points) into a random field
that has a continuous and piecewise smooth sample path.
In Section \ref{sec:detection}, 
we analyze the data of \citet{Mizuta-etal10}.
They first screened the candidates of loci by analyzing
datasets from two $\mathrm{F}_2$ populations and 
reciprocal backcross (BC) populations,
and finally succeeded in isolating causal genes of a BDM pair by positional cloning.
We 
examine their data, and confirm that their genetic finding
about the BDM pair is significant from the viewpoint of
multiple testing procedures.
The proofs of Proposition \ref{prop:chi-square-field}, which describes the
asymptotic correlation structure of the chi-square statistics
for detecting interactive pairs, and
the tail probability formulas in Theorems \ref{thm:renewal} and \ref{thm:tube}
are given in 
Section \ref{sec:proofs}. 

\subsection{Scan statistics for the detection of interactive loci pairs}
\label{subsec:scan}

In this subsection, we explain the scan statistic for detecting BDM pairs
and its asymptotic joint distribution for the case of the $\mathrm{F}_2$
population dealt with by \citet{Mizuta-etal10}.

We focus on the number of $\mathrm{F}_2$ 
individuals that avoided such a fatal event and grew up.
Each locus of an individual in the $\mathrm{F}_2$ population produced by
two strains A and B has the genotypes AA, BB, and AB.
Abbreviating them to A, B, and H, respectively, the genotypes
of loci 1 and 2 are cross-classified in Table \ref{tab:crosstable}.
\begin{table}[htbp]
\caption{Cross table of genotypes in two loci ($\mathrm{F}_2$)}
\label{tab:crosstable}
\begin{center}
\begin{tabular}{c|ccc}
\thline
locus 1 $\backslash$ locus 2 & A & B & H \\
\hline
A  & $n_\mathrm{AA}$ & $n_\mathrm{AB}$ & $n_\mathrm{AH}$ \\
B  & $n_\mathrm{BA}$ & $n_\mathrm{BB}$ & $n_\mathrm{BH}$ \\
H  & $n_\mathrm{HA}$ & $n_\mathrm{HB}$ & $n_\mathrm{HH}$ \\
\thline
\end{tabular}
\end{center}
\end{table}
If this table shows some discrepancy against the independence of rows and columns,
then the lack of individuals (sterility) is assumed to have happened
when the loci pair has particular genotypes.  
Noting this, \citet{Mizuta-etal10} used
the chi-square statistics for independence (Pearson's chi-square statistics)
as scan statistics for detection.
Similar scan statistics are used by \citet{Kao-etal10}
in an $\mathrm{F}_1$ spore population from an inter-species cross of yeast.

Let $T_{c_1 c_2}(j_1,j_2)$ ($c_1<c_2$) be the chi-square statistic
calculated from the pair of the marker $j_1$ on chromosome $c_1$
and the marker $j_2$ on chromosome $c_2$.
The multiplicity-adjusted $p$-value can be obtained from the upper probability
of the maximum of all chi-square statistics
$\max_{c_1<c_2} \max_{j_1,j_2} T_{c_1 c_2}(j_1,j_2)$
under the null hypothesis $H_0$ that a BDM pair does not exist.
The distribution of each statistic $T_{c_1 c_2}(j_1,j_2)$ is
approximated as the chi-square distribution with 4 degrees of freedom
when the number $n$ of individuals is large.
However, these statistics are not independent
and are highly correlated because of the linkage.
Under the assumption of Haldane's model
 (see, e.g., \citet{Siegmund-Yakir07}, Section 5.6),
which is the most standard model for linkage, the joint distribution under
the null hypothesis $H_0$ is described in
Proposition \ref{prop:chi-square-field} below.
The proof is given in 
Section \ref{subsec:proof-corr}. 
\begin{prop}
\label{prop:chi-square-field}
(a) Let $d_{1 j_1}$ (M: Morgan) be locations of markers $j_1\,(=1,\ldots,m_1)$
 on a chromosome (chromosome $1$, say).
Let $d_{2 j_2}$ be locations of markers $j_2\,(=1,\ldots,m_2)$
 on another chromosome (chromosome $2$, say).
Under the null hypothesis that a BDM pair does not exist,
as the total sample size $n$ goes to infinity, convergence in distribution
\begin{equation}
\label{convergence}
T_{12}(j_1,j_2) \Rightarrow
 Z_1(j_1,j_2)^2 + Z_2(j_1,j_2)^2 + Z_3(j_1,j_2)^2 + Z_4(j_1,j_2)^2
 \quad (n\to\infty)
\end{equation}
holds jointly for all $(j_1,j_2)$,
where 
$Z_1,\ldots,Z_4$ are independent,
and for each $k$, the $Z_k(i_1,i_2)$'s are distributed
according to the multivariate normal distribution
with a marginal mean $0$, a variance $1$, and the following covariance structure:
\begin{equation}
\label{cov-structure}
 \Cov(Z_k(i_1,i_2),Z_k(j_1,j_2))
= e^{-\rho_{k1}|d_{1 i_1}-d_{1 j_1}|} \times
  e^{-\rho_{k2}|d_{2 i_2}-d_{2 j_2}|}
\end{equation}
with
\begin{equation}
\label{rhok12}
 (\rho_{k1},\rho_{k2}) = \begin{cases}
 (2,2) \ (k=1), & (2,4) \ (k=2), \\
 (4,2) \ (k=3), & (4,4) \ (k=4). \end{cases}
\end{equation}

(b) Under the null hypothesis that a BDM pair does not exist,
$T_{c_1 c_2}$ and $T_{c_1' c_2'}$ are asymptotically independently
distributed unless $(c_1,c_2)=(c_1',c_2')$.
\end{prop}

This proposition does not tell us about marker pairs
belonging to the same chromosome.
When two markers are located on the same chromosome,
the linkage affects the independence of the rows and columns
in Table \ref{tab:crosstable}, and the chi-square statistic simply measures
the effect of the linkage directly.
Because this is irrelevant to the reproductive isolation,
we ignore such pairs.

Based on the asymptotic distribution given by
Proposition \ref{prop:chi-square-field}, we can evaluate
the multiplicity-adjusted $p$-value (see (\ref{pvalue})).
Actually, in our genetic application,
the sample size $n$ is large enough (more than 100, at least), and this
asymptotic approximation works well (see, Section \ref{subsec:numerical2}).
In this context, calculation of the upper probability of the maximum
of a chi-square random field on lattice points is crucial. 
The primary theoretical purpose of this paper is to provide approximate
formulas for upper tail probability in a more general setting. 

\subsection{Other examples}
\label{subsec:other}

The covariance structure in Proposition \ref{prop:chi-square-field}
also appears in other scan statistics.
We illustrate two examples briefly.

The first example is the detection of epistasis 
in quantitative trait loci (QTL) analysis.
In QTL analysis for $\mathrm{F}_2$ population,
phenotype $y$ and genotypes $z_j$ are observed for each individual,
where $j$ is the index of markers, and $z_j$ takes the values A, B, and H.
The following is a simple model of QTL analysis incorporating the effects of
epistasis between a loci pair $(j_1,j_2)$:
\[
y=\mu
  + \sum_j (\alpha_j  v_j + \beta_j w_j)
  + \gamma_1 v_{j_1} v_{j_2} + \gamma_2 v_{j_1} w_{j_2}
  + \gamma_3 w_{j_1} v_{j_2} + \gamma_4 w_{j_1} w_{j_2} + \varepsilon,
\]
where
$v_j=1$ ($z_j=\mathrm{A}$), $=0$ ($z_j=\mathrm{H}$), $=-1$ ($z_j=\mathrm{B}$),
$w_j=1$ ($z_j=\mathrm{A},\mathrm{B}$), $=-1$ ($z_j=\mathrm{H}$),
and $\varepsilon$ is a Gaussian measurement error.
The parameters $\gamma_1,\ldots,\gamma_4$ represent the epistasis.
For identifying the loci pair $(j_1,j_2)$,
the scan statistic $U(j_1,j_2)$ defined as the likelihood ratio test (LRT)
statistic for testing the null hypothesis of no epistasis
$\gamma_1,\ldots,\gamma_4=0$ is used.
It is shown that the asymptotic joint distribution of $\{U(j_1,j_2)\}$ is the
same as that of $\{T(j_1,j_2)\}$ in Proposition \ref{prop:chi-square-field}
when $j_1$ and $j_2$ are on different chromosomes, and the
multiplicity-adjusted $p$-value can be obtained similarly.

The second example is the detection of a change-point
in two-way ordered categorical data.
For the cell probability $\{p_{ij}\}_{a\times b}$, \citet{Hirotsu97} assumed
a log-linear model with a change-point at $(i_0,j_0)$:
\[
 \log p_{ij} = \alpha_i + \beta_j + \gamma \1(i\le i_0,j\le j_0),
\]
where $\1(\cdot)$ is the indicator function,
and define a scan statistic $V(i_0,j_0)$ as the LRT statistic
for testing $\gamma=0$.
Under the null hypothesis,
$\{V(i_0,j_0)\}_{i_0=1,\ldots,a,\,j_0=1,\ldots,b}$ is asymptotically
equivalent to $\{Z_1(j_1,j_2)^2\}$ in Proposition \ref{prop:chi-square-field}
with $d_{1j}=\log\frac{P_j}{1-P_j}$,
$d_{2j}=\log\frac{Q_j}{1-Q_j}$,
$P_i=\sum_{k=1}^i\sum_{l=1}^b p_{kl}$,
$Q_j=\sum_{k=1}^a\sum_{l=1}^j p_{kl}$,
and multiplicity-adjusted $p$-value can be obtained in our framework.

\section{Approximate tail probabilities}
\label{sec:maximum}

\subsection{Chi-square random fields restricted on lattice points}
\label{subsec:problem}

In this section, as a generalization of the random field referred to in
Proposition \ref{prop:chi-square-field}, we define a chi-square random field
on a multi-dimensional index set with a direct-product type covariance
structure such as (\ref{cov-structure}), and consider the distribution of
its maximum over a multi-dimensional lattice points.

For $k=1,\ldots,m$,
let us consider a real-valued continuous Gaussian random field
on $\R^p$ that has the following moment structure:
\[
 E[Z_k(t)]=0, \quad V[Z_k(t)]=1, \quad \Cov(Z_k(t),Z_k(t')) = R_k(t-t'),
\]
where for $h=(h_1,\ldots,h_p)$,
\begin{equation}
\label{R}
 R_k(h)=\prod_{i=1}^p R_{ki}(h_i), \quad R_{ki}(h_i)=1-\rho_{ki}|h_i|+o(|h_i|)\ \ %
\mbox{as $h_i\to 0$},
\end{equation}
and $\rho_{ki}$ is a positive constant.
In particular, when $R_{ki}(h_i)=e^{-\rho_{ki}|h_i|}$,
this expression represents the direct-product covariance structure of the stationary
Ornstein-Uhlenbeck process.
$Z_1,\ldots,Z_m$ are assumed to be independent.
Moreover, define
\begin{equation}
\label{Y}
 Z(t)=(Z_1(t),\ldots,Z_m(t)), \qquad
 Y(t) = \Vert Z(t)\Vert = \sqrt{\sum_{k=1}^m Z_k(t)^2}.
\end{equation}
$Y(t)^2$, $t=(t_1,\ldots,t_p)\in\R^p$ is a chi-square random field
whose marginal distribution is the chi-square distribution
with $m$ degrees of freedom.

For $i=1,\ldots,p$, let $0=d_{i0}<d_{i1}<\cdots <d_{i n_i}$ be distinct points,
and let $T_i=\{d_{i0}\,(=0),d_{i1},\ldots,d_{i n_i}\}$.
Define a $p$-dimensional unequally spaced lattice point set
\[
 T = T_1\times\cdots\times T_p \subset \R^p.
\]
In this section, we provide an approximate formula for the tail probability
of the maximum of the chi-square random field $Y$
restricted on the discrete set $T$:
\begin{equation}
\label{maxchi}
 P\Bigl( \max_{t\in T} Y(t) \ge b \Bigr) \quad \mbox{as $b\to\infty$}.
\end{equation}

\subsection{Approximations based on nonlinear renewal theory}
\label{subsec:renewal}

In this subsection, we study large-deviation approximations for
the distribution of the maximum (\ref{maxchi})
in the framework of the nonlinear renewal theory devised by
\citet{Woodroofe82} and \citet{Siegmund88}.
The outline of this method is that we first prove that
$\max_{t\in T} Y(t)$ can be approximated by the maximum of a
suitably defined random walk when $Y$ is large and
the spacing of lattice is small.
We then evaluate the distribution of its maximum
with the help of sequential analysis.

A drawback of the method is that the index set $T$ must be
an equally spaced lattice point set.  That is,
for all $i$, the points $d_{i0}<\cdots<d_{i n_i}$ belonging to $T_i$ are
assumed to be equally spaced as
\[
 d_{i1}-d_{i0} = \cdots = d_{i n_i}-d_{i n_i-1} \, (= D_i,\ \mbox{say}).
\]
If the spaces are not equal, the random walk in the limit
does not approach the sum of identical distributions,
and hence one cannot utilize the reproductivity in the sequential analysis.
However, as we show in Section \ref{subsec:numerical}, in
typical settings
for genome analysis,
the upper probability for the maximum on unequally spaced lattice points is 
bounded above by that for the maximum on the equally spaced lattice
 (i.e., the latter gives a conservative bound for the former),
and the difference between them is not substantial.

Define a bounded rectangle in $\R^p$ by
\[
 \widetilde T=\widetilde T_1\times\cdots\times\widetilde T_p \subset\R^p, \quad \widetilde T_i=[0,d_{i n_i}].
\]
For
\begin{equation}
\label{multiindex}
 j=(j_1,\ldots,j_p)\in \Z^p, \quad D=(D_1,\ldots,D_p)\in \R^p,
\end{equation}
we write
$ j D = (j_1 D_1,\ldots,j_p D_p) $.
Our problem is to approximate the distribution of the maximum on
$p$-dimensional lattice points whose spacing in the $i$th coordinate is $D_i$
as follows:
\[
 P\Bigl( \max_{j\in J} Y(j D) \ge b \Bigr),
 \quad J= \Bigl\{ j\in\Z^m \mid j D \in \widetilde T \Bigr\},
 \quad \mbox{as $b\to\infty$}.
\]

By using the approach of nonlinear renewal theory, we can obtain the following
formula.  The proof is given in 
Section \ref{subsec:proof-renewal}. 

\begin{thm}
As $b\to\infty$, $D_i\to 0$ such that $b\sqrt{D_i}\to c_i\in (0,\infty)$,
 $i=1,\ldots,p$,
\label{thm:renewal}
\begin{equation}
\label{renewal}
P\Bigl( \max_{j\in J} Y(j D) \ge b \Bigr)
\,\sim\,%
 \frac{|\widetilde T|}{(2\pi)^{m/2}} b^{m+2p-2} e^{-b^2/2}
  \int_{\S^{m-1}} \prod_{i=1}^p \bar\rho_i \nu(b\sqrt{2\bar\rho_i D_i}) \, du,
\end{equation}
where $du$ is the volume element of the unit sphere $\S^{m-1}$ in $\R^m$
at $u = (u_1,\ldots,u_m) \in \S^{m-1}$,
\begin{equation}
\label{rhobar}
 \bar\rho_i = \bar\rho_i(u) = \sum_{k=1}^m u_k^2 \rho_{ki},
\end{equation}
$|\widetilde T|$ is the Lebesgue measure of $\widetilde T$, and
\[
\nu(x) = \begin{cases}
 2 x^{-2} \exp\Bigl\{ -2 {\sum}_{n=1}^\infty n^{-1}
 \Phi\Bigl(-\frac{1}{2}x\sqrt{n}\Bigr) \Bigr\} & (x>0), \\
 1 & (x=0) \end{cases}
\]
with $\Phi(\cdot)$ the cumulative distribution function of
the standard normal distribution.
\end{thm}

It is reported that the asymptotic setting where
 $D_i=O(b^{-2})$ as $b\to\infty$ assumed in
Theorem \ref{thm:renewal} leads to good approximation formulas
in QTL analysis when makers are dense
(\citet{Dupuis-Siegmund99}, \citet{Siegmund04}, and \citet{Siegmund-Yakir07}).

\begin{rem}
\label{rem:nu}
The function $\nu(x)$ can be conveniently approximated by the following:
\begin{equation}
\label{nu}
 \nu(x) \approx \frac{(2/x)(\Phi(x/2)-1/2)}{(x/2)\Phi(x/2)-\phi(x/2)},
\end{equation}
where $\phi(\cdot)$ is the density function of the standard normal distribution (\citet{Siegmund-Yakir07}). 
We use this in numerical calculations presented in Section \ref{subsec:numerical}.
\end{rem}

\begin{rem}
\label{rem:piterbarg}
The upper tail probability of the maximum of a continuous chi random field
$Y$ over a continuous set $\widetilde T$ can be obtained by following
\citet{Piterbarg96}, Corollary 7.1 
as follows:
\begin{equation}
\label{piterbarg}
 P\Bigl( \max_{t\in \widetilde T} Y(t) \ge b \Bigr)
 \,\sim\,\frac{|\widetilde T|}{(2\pi)^{m/2}}
 b^{m+2p-2} e^{-b^2/2}
 \int_{\S^{m-1}} \prod_{i=1}^p \bar \rho_i(u) \, du
 \quad (b\to\infty).
\end{equation}
This is coincident with the right-hand side of (\ref{renewal}) with $c_i=0$.
Since $\max_{t\in T} Y(t) \le \max_{t\in \widetilde T} Y(t)$,
 (\ref{piterbarg}) is an asymptotic upper bound for (\ref{maxchi}).
This can be confirmed directly from the fact $\nu(x)\le 1$.
\end{rem}

\begin{rem}
\label{rem:bonferroni}
The Bonferroni bound of the left-hand side of (\ref{renewal}) is
\[
P\Bigl( \max_{j\in J} Y(j D) \ge b \Bigr)
\,\le\,%
 |J|\,P\bigr(\chi^2_m\ge b^2\bigl),
\]
where $\chi^2_m$ is a chi-square random variable with $m$ degrees of freedom.
As $b\to\infty$,
this Bonferroni bound is asymptotically evaluated as
\begin{equation}
\label{bonferroni-asympt}
 \frac{|\widetilde T|}{(2\pi)^{m/2}} b^{m+2p-2} e^{-b^2/2}
 \frac{1}{\prod_{i=1}^p (b^2 D_i)} \int_{\S^{m-1}} du.
\end{equation}
Here, we used $|J|=|\tilde T|/\prod_{i=1}^p D_i$,
$P(\chi^2_m\ge b^2) \sim b^{m-2} e^{-b^2/2}/2^{m/2-1}\Gamma(\frac{m}{2})$
and $\int_{\S^{m-1}} du = 2\pi^{m/2}/\Gamma(\frac{m}{2})$.
The right-hand side of (\ref{renewal}) is actually bounded above by
 (\ref{bonferroni-asympt}) because of $\nu(x)\le 2 x^{-2}$.
\end{rem}

\subsection{Approximations based on the volume-of-tube method}
\label{subsec:tube}

In this subsection, we provide a conservative bound for the distribution of
the maximum of a chi-square random field (\ref{maxchi}) by adopting
an integral-geometric approach referred to as the volume-of-tube method or
the Euler characteristic heuristic.

The volume-of-tube method approximates the distribution of the maximum of
a Gaussian random field that has a continuous and piecewise smooth sample path.
It is particularly useful when the marginal distribution
(with a fixed index) is standard normal $N(0,1)$.
 (See,  
\citet{Sun93},
\citet{Kuriki-Takemura01,Kuriki-Takemura09}, \citet{Takemura-Kuriki02}, and
\citet{Adler-Taylor07}%
.)
In order to apply the volume-of-tube method to our problem, we need to describe
our problem in terms of a Gaussian random field
with a continuous and piecewise smooth sample path.

First,
we modify the Gaussian random field $Z_k$ on a discrete set $T$
to define a Gaussian random field $\widetilde Z_k$ on a continuous set
$\widetilde T$ that has the following properties:
\begin{itemize}
\item[(a)]
$Z_k(t) = \widetilde Z_k(t)$ (if $t\in T$).
\item[(b)]
As a function of $t\in \widetilde T$, $\widetilde Z_k(t)$ is continuous and
piecewise smooth.
\end{itemize}
Note that continuous processes with the covariance structures given by (\ref{R})
do not satisfy (b).
This is because the covariance function is not differentiable at $h=0$,
and hence the sample path is not differentiable everywhere.

Define a chi random field on the index set $\widetilde T$ by
\[
 \widetilde Y(t) = \sqrt{\sum_{k=1}^m \widetilde Z_k(t)^2}.
\]
In addition, define a Gaussian random field on the index set
$\widetilde T\times\S^{m-1}$ by
\[
 \widetilde X(t,u) = \sum_{k=1}^m u_k \widetilde Z_k(t),
 \quad u=(u_1,\ldots,u_m)\in\S^{m-1}.
\]
Since $Y(t)=\widetilde Y(t)=\max_{u\in\S^{m-1}} \widetilde X(t,u)$
for $t\in T$, we can use the upper probability of
$\max_{t\in \widetilde T} Y(t)
 = \max_{(t,u)\in \widetilde T\times\S^{m-1}} \widetilde X(t,u)$
as a conservative bound for that of $\max_{t\in T} Y(t)$.
Note that $\widetilde X(t,u)$ with $(t,u)$ fixed has
a standard normal distribution.

Under the volume-of-tube method, the index set $\widetilde T\times\S^{m-1}$
is regarded as a Riemannian manifold endowed with a metric of
\begin{equation}
\label{metric}
 g(t,u) = \Cov\bigl(\nabla_{(t,u)}\widetilde X(t,u),\nabla_{(t,u)}\widetilde X(t,u)\bigl)
\end{equation}
at $(t,u)$.
When a positive definite metric can be defined by (\ref{metric}),
approximate tail probability formulas can be obtained as asymptotic
expansions involving geometric invariants measured by this metric.
However, even when the index set contains singularities
where the metric is not properly defined,
if the volume $\Vol(\widetilde T\times\S^{m-1})$ of the index set can only be
evaluated by integrals over regular sets,
the leading-term formula given below applies (\citet{Takemura-Kuriki03}).
Note that the dimension of the index set is
$\dim(\widetilde T\times\S^{m-1})=p+m-1$.
\begin{align}
 P\biggl(\max_{t\in \widetilde T} \widetilde Y(t) & \ge b\biggr)
 \,=\,P\biggl(\max_{(t,u)\in \widetilde T\times\S^{m-1}}\widetilde X(t,u)
              \ge b\biggr)  \nonumber \\
 \sim& \, \Vol\bigl(\widetilde T\times\S^{m-1}\bigr) \cdot
   \frac{2}{(2\pi)^{(p+m)/2}} \, b^{p+m-2} e^{-b^2/2} \quad (b\to\infty).
\label{tube0}
\end{align}

There is no unique way of constructing a $\widetilde Z_k$
satisfying (a) and (b) from $Z_k$.
We construct $\widetilde Z_k$ by undertaking the following steps.
\begin{itemize}
\item[(i)]
Dissect the $p$-dimensional rectangle whose vertices are flanking lattice
points of $T$,
\[
 [d_{1 j_1-1},d_{1 j_1}]\times\cdots\times[d_{p j_p-1},d_{p j_p}],
\]
into $p!$ simplices.
\item[(ii)]
For each simplex, define $\widetilde Z_k$ over the simplex
by linearly interpolating the values of $Z_k$ at vertices
and multiplying by a scalar so that the variance of $\widetilde Z_k$
at each point of the simplex is 1.
\end{itemize}
Details of the proof of the next theorem and details of how to construct
$\widetilde Z_k$ are given in 
Section \ref{subsec:proof-tube}. 

\begin{thm}
\label{thm:tube}
Let $D_{ij}=d_{i j}-d_{i j-1}$.
As $b\to\infty$ and $\max D_{ij}\to 0$,
\begin{equation}
\label{tube}
P\Bigl( \max_{t\in T} Y(t) \ge b \Bigr)
 \,\le\,  P\biggl( \max_{t\in \widetilde T} \widetilde Y(t) \ge b \biggr)
 \,\sim\, \frac{2 V}{(2\pi)^{(m+p)/2}} b^{m+p-2} e^{-b^2/2},
\end{equation}
where
\[
 V = 2^{p/2}\,\prod_{i=1}^p \sum_{j=1}^{n_i} \sqrt{D_{ij}}
 \int_{\S^{m-1}} \prod_{i=1}^p 
 \sqrt{\bar\rho_i(u)} \,du,
\]
and $\bar\rho_i(u)$ is defined in (\ref{rhobar}).
In addition, $du$ is the volume element of $\S^{m-1}$ at $u$.
\end{thm}

\begin{rem}
The polynomial factor $b^{m+p-2}$ in (\ref{tube}) is smaller than
$b^{m+2p-2}$ in (\ref{renewal}) and (\ref{piterbarg}).
However, this does not imply that
 (\ref{tube}) is a better bound than (\ref{renewal}).
As $\max D_{ij}\to 0$, $\sum_{j=1}^{n_i} D_{ij}=O(1)$,  
$\sum_{j=1}^{n_i} \sqrt{D_{ij}}\ge \sum_{j=1}^{n_i} D_{ij}/\sqrt{\max D_{ij}}\to \infty$, and hence $V\to\infty$.
$V$ is not of constant order.
\end{rem}

\citet{Ninomiya04} provided a conservative bound for the upper probability
of the maximum of a Gaussian random field on a 2-dimensional lattice
with a product-type covariance structure (\ref{R})
in detecting a change-point in two-way ordered categorical data.
\citet{Rebai-etal94} also applied the volume-of-tube method to linkage analysis.
He computed thresholds for the maximum log odds (LOD) score
in the interval mapping method
by using Rice's formula, which is essentially equivalent to the volume-of-tube method.

\subsection{Numerical comparisons of proposed formulas}
\label{subsec:numerical}

This and succeeding subsections are devoted to numerical studies.
In this subsection, we make numerical comparisons of three approximations:
the formula based on nonlinear renewal theory (Theorem \ref{thm:renewal});
the conservative bound based on continuous processes (Remark \ref{rem:piterbarg});
and the conservative bound based on the volume-of-tube method (Theorem \ref{thm:tube}).
The Bonferroni method (Remark \ref{rem:bonferroni}) is also included
as a reference.
Mindful of the problem of detecting the interactive loci pairs (BDM pairs),
as explained in Section \ref{sec:intro},
we set the parameters as follows:
The dimension of the index set is $p=2$,
the chi-square degrees of freedom is $m=4$
and $1$,
$(\rho_{k1},\rho_{k2})$ ($k=1,2,3,4$) are in (\ref{rhok12}),
$n_1=n_2=50,\,100,\,200$,
$D_{1 j}=D_{2 j}\equiv 0.2/100,\,1/100,\,5/100$ (equally spaced),
$(D_{1 j})_{j\ge 1}=(D_{2 j})_{j\ge 1}=(0.5,1,0.5,1,3,0.5,1,0.5,1,1,\ldots)/100$
(repeat the cycle with period 10) (pattern I),
$(D_{1 j})_{j\ge 1}=(D_{2 j})_{j\ge 1}=(0.5,0.5,3,0.5,0.5,\ldots)/100$
(repeat the cycle with period 5) (pattern II),
$\widetilde T=[0,1]^2$.
Note that the length $1/100$ corresponds to 1cM on a chromosome.

Let $U=(U_1,\ldots,U_m)$ be a random vector with a
uniform distribution on the unit sphere $\S^{m-1}$ in $\R^m$.
An integral over $\S^{m-1}$ with respect to the volume element $du$
can be replaced by the expectation
$\int_{\S^{m-1}} f(u) du = \Vol(\S^{m-1}) \, E[f(U)]$,
$\Vol(\S^{m-1})=2\pi^{m/2}/\Gamma(m/2)$.
In particular, we use the following
for $m=4$
and $(\rho_{k1},\rho_{k2})$ given in (\ref{rhok12}):
\begin{align*}
& E\biggl[\prod_{i=1}^2 \bar\rho_i(U)\biggr]
 = \frac{\prod_{i=1}^2(\sum_{k=1}^m \rho_{ki})
        +2\sum_{k=1}^m \rho_{k1}\rho_{k2}}{m(m+2)} = 9, \\
& E\biggl[\prod_{i=1}^2 \sqrt{\bar\rho_i(U)}\biggr] \doteq 2.971.
\end{align*}

Moreover, we use the approximation (\ref{nu}) in calculating
the special function $\nu(x)$.

\begin{figure}[h]
\begin{center}
 \hspace*{-30mm}
 \rotatebox{-90}{\scalebox{0.6}{\includegraphics{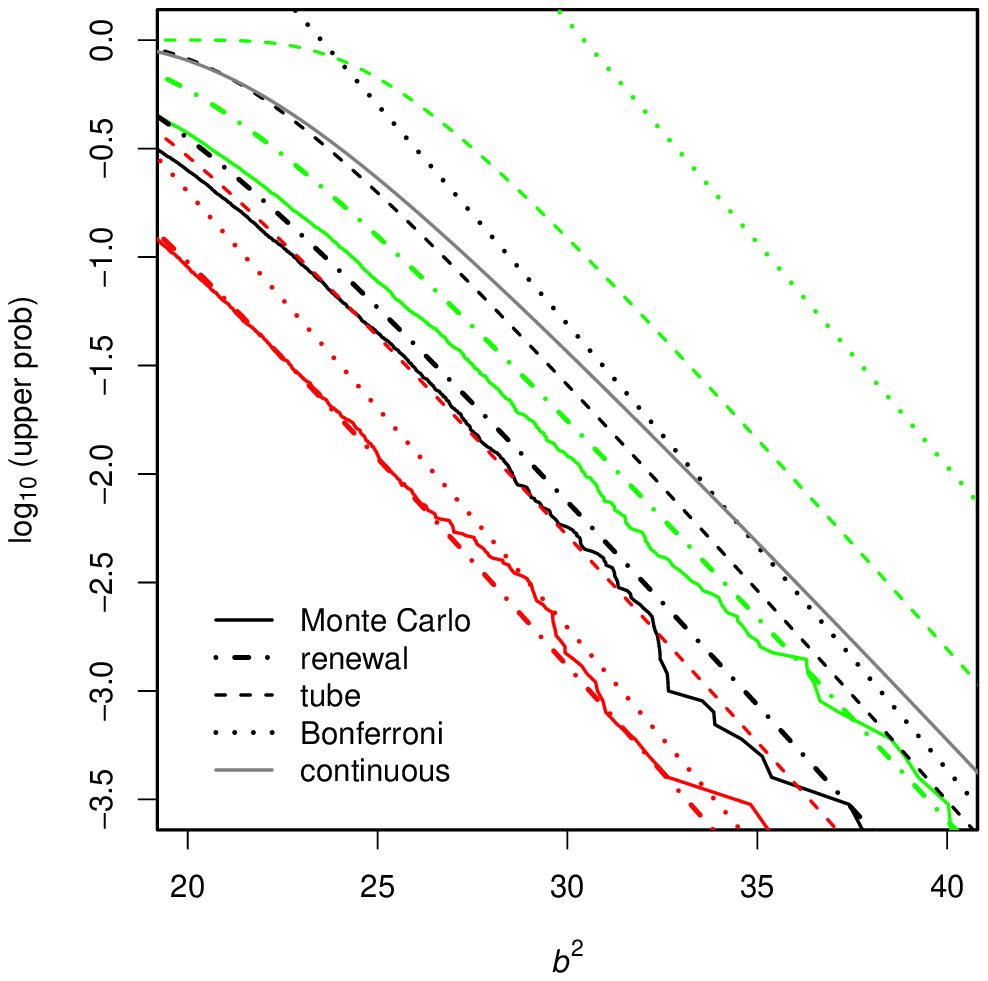}}}
\end{center}
\caption{
Comparisons of upper probability formulas (equally spaced case).
\newline
Degrees of freedom $m=4$, $\widetilde T=[0,1]^2$,
$D_{ij}\equiv 0.05$ (red), $0.01$ (black), $0.002$ (green).
Continuous approximation is in gray.
Monte Carlo simulations were based on 10,000 iterations.
}
\label{fig:ou2-delta}
\end{figure}

\begin{figure}[h]
\begin{center}
 \hspace*{-30mm}
\rotatebox{-90}{\scalebox{0.6}{\includegraphics{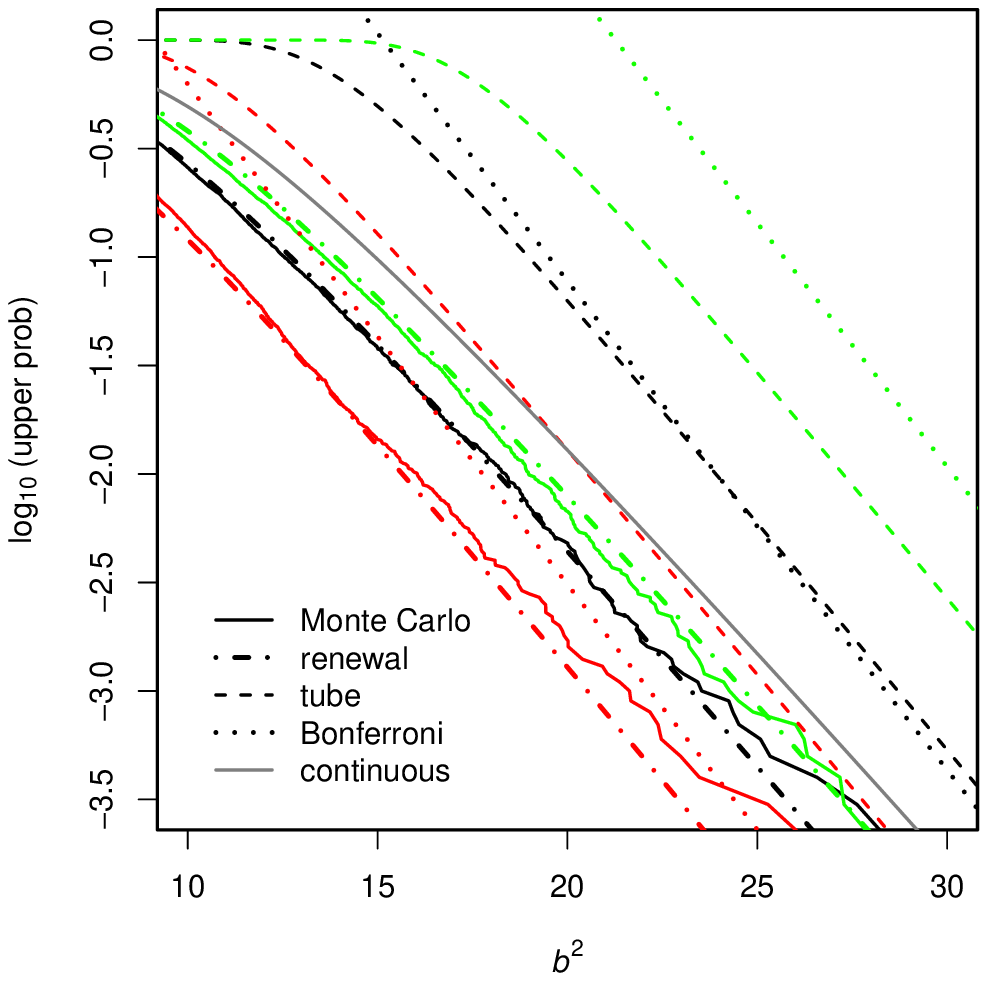}}}
\end{center}
\caption{
Comparisons of upper probability formulas (equally spaced case).
\newline
Degrees of freedom $m=1$, $\widetilde T=[0,1]^2$,
$D_{ij}\equiv 0.05$ (red), $0.01$ (black), $0.002$ (green).
Continuous approximation is in gray.
Monte Carlo simulations were based on 10,000 iterations.
}
\label{fig:ou2-df1}
\end{figure}

\begin{figure}[h]
\begin{center}
 \hspace*{-30mm}
 \rotatebox{-90}{\scalebox{0.6}{\includegraphics{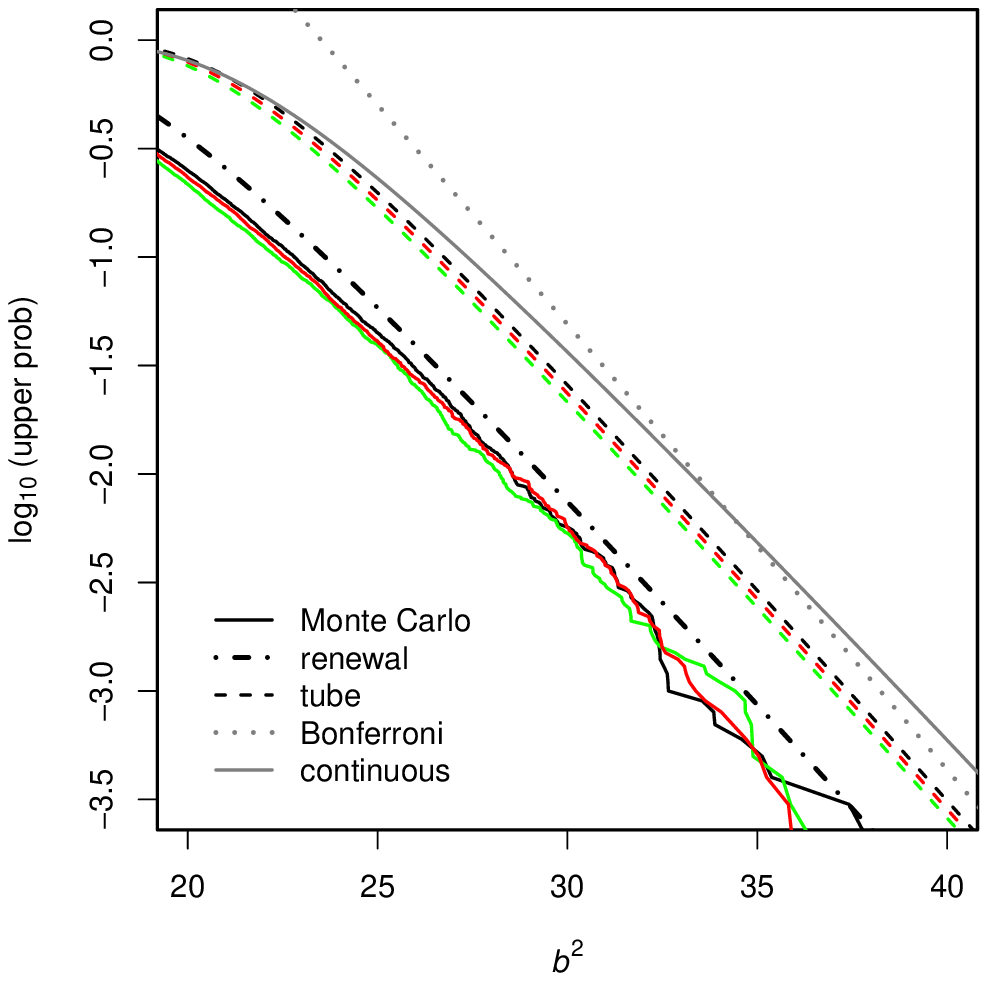}}}
\end{center}
\caption{
Comparisons of upper probability formulas (unequally spaced case).
\newline
Degrees of freedom $m=4$, $\widetilde T=[0,1]^2$,
$D_{ij}\equiv 0.01$ (black), 
pattern I: $(D_{ij})_{j\ge 1}=(0.5,1,0.5,1,3,0.5,1,0.5,1,1,\ldots)/100$ (red),
pattern II: $(D_{ij})_{j\ge 1}=(0.5,0.5,3,0.5,0.5,\ldots)/100$ (green).
Continuous approximation and the Bonferroni bound are is in gray.
Monte Carlo simulations were based on 10,000 iterations.
}
\label{fig:ou2-unequal}
\end{figure}

Figures \ref{fig:ou2-delta}--\ref{fig:ou2-unequal} illustrate
the comparisons among three approximate formulas
as well as empirical distributions of Monte Carlo simulations
with 10,000 iterations
for the probability $P\bigl(\max_{t\in T} Y(t)^2\ge b^2\bigr)$.
Random numbers are generated from the following spatial autoregressive model:
For $k=1,\ldots,m$, $i=0,1,\ldots,n_1\,(=100)$, $j=0,1,\ldots,n_2\,(=100)$,
let $\varepsilon_k(i,j)$ be independent standard normal distributed random variables.
Generate $Z_k(i,j)$ sequentially according to
\begin{equation}
\label{AR}
\begin{cases}
\displaystyle
Z_k(0,0) \,=\,
 \varepsilon_k(0,0), \\
\displaystyle
Z_k(i,0) \,=\,
 \alpha_k(i) Z_k(i-1,0)
 + \sqrt{1-\alpha_k(i)^2} \, \varepsilon_k(i,0) & (i\ge 1), \\
\displaystyle
Z_k(0,j) \,=\,
 \beta_k(j) Z_k(0,j-1)
 + \sqrt{1-\beta_k(j)^2} \, \varepsilon_k(0,j) & (j\ge 1), \\
Z_k(i,j) \,=\,
 \alpha_k(i) Z_k(i-1,j) + \beta_k(j) Z_k(i,j-1) \\
 \qquad\qquad\quad
\displaystyle
- \alpha_k(i) \beta_k(j) Z_k(i-1,j-1) \\
 \qquad\qquad\quad
\displaystyle
 + \sqrt{1-\alpha_k(i)^2} \sqrt{1-\beta_k(j)^2} \, \varepsilon_k(i,j)
 & (i,j\ge 1),
\end{cases}
\end{equation}
where
\[
 \alpha_k(i) = e^{-\rho_{k1} D_{1i}}, \quad \beta_k(j) = e^{-\rho_{k2} D_{2j}}.
\]
Then,
\[
 \max_{i,j\ge 0} Y(i,j)^2 = \max_{i,j\ge 0} \, \sum_{k=1}^4 Z_k(i,j)^2
\]
is obtained.
In these figures,
the transformed upper probabilities
of the three approximate formulas by using the transformation $x\mapsto 1-e^{-x}$ are depicted.
This map is adopted by \citet{Dupuis-Siegmund99}, (9),
to restrict the maximum $p$-value to less than 1
without altering the asymptotic behaviors of the tail probabilities.

Figures \ref{fig:ou2-delta} and \ref{fig:ou2-df1} show that
the formula based on nonlinear renewal theory approximates the tail probabilities well
in wide ranges of the marker spacing, length of chromosomes.
In particular, the case where the degree $m$ of freedom is $1$ shows
greater accuracy than when $m=4$.
We conclude that the asymptotic setting where $D_i=O(b^{-2})$ ($b\to\infty$)
assumed in Theorem \ref{thm:renewal} fit to our genetic applications
where the marker spacings is fairly small.
On the other hand,
the formulas based on the volume-of-tube method and the continuous process yield
upper bounds for the upper probabilities.
Neither of these two methods is superior to the other.
The Bonferroni method is always most conservative.

Figure \ref{fig:ou2-unequal} shows that the statistics for
unequally
spaced sampling are
slightly below those for equally spaced sampling.
This suggests that
the formulas for equally spaced lattice lead to conservative $p$-value estimators
when the sampling spaces are unequal.

\subsection{Adequacy of asymptotic approximation}
\label{subsec:numerical2}

\begin{figure}[t]
\begin{center}
 \hspace*{-30mm}
 \rotatebox{-90}{\scalebox{0.6}{\includegraphics{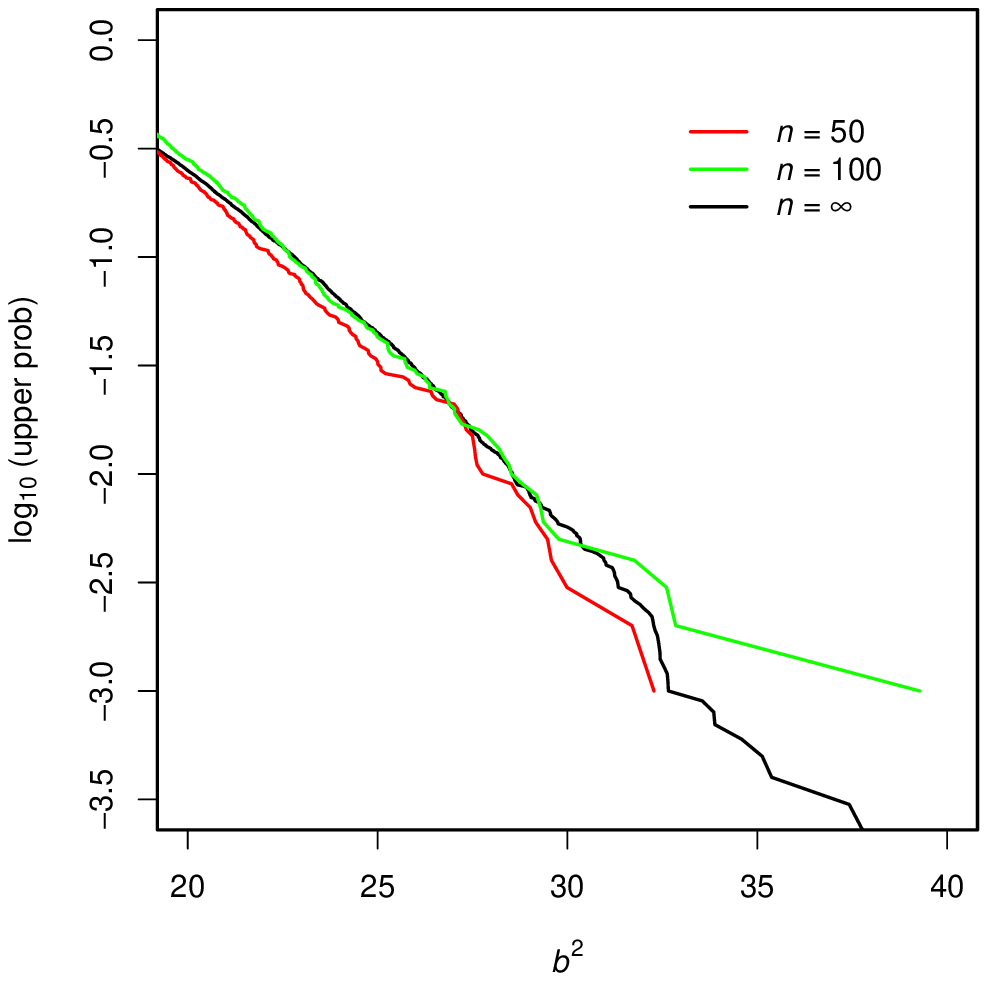}}}
\end{center}
\caption{
Tail probabilities when $n$ is finite and infinite.
\newline
Degrees of freedom $m=4$, $\widetilde T=[0,1]^2$, $D_{ij}\equiv 0.01$.
Numbers of iterations were 1,000 ($n<\infty$), 10,000 ($n=\infty$).
}
\label{fig:ou2-finite}
\end{figure}

Throughout the paper, our arguments rely on the asymptotic approximation
of Pearson's statistics to chi-square statistics.
For a single contingency table, it is said that this approximation works well
practically if expected cell frequencies are greater than 5
(\citet{Agresti02}, Section 3.2.1).
The sample size in our application is large enough, and 
this criterion holds for each loci pair table in Table \ref{tab:crosstable}.
However, we need to be careful since
we are coping with a joint distribution of many tables.
Figure \ref{fig:ou2-finite} depicts the upper probabilities of the statistics
in both cases where the sample size $n$ is finite and infinite
by Monte Carlo simulations.
The setting of experiments is the same as in Figure \ref{fig:ou2-delta}
with $D_{ij}\equiv 0.01$.
The curve for $n=\infty$ is the same as in Figure \ref{fig:ou2-delta}.
The curves for $n<\infty$ are estimated by Monte Carlo simulations
with 1,000 replications.
For the case $n<\infty$, we first generate the sequences
of genotypes $\epsilon_i^{(t)}$, $\delta_i^{(t)}$,
$\widetilde\epsilon_j^{(t)}$, $\widetilde\delta_j^{(t)}$ by means of
Markov property (\ref{markov}), calculate $T_{ij}$ by (\ref{3x3}),
and take the maximum $\max_{i,j} T_{ij}$.
Figure \ref{fig:ou2-finite} suggests that asymptotic approximation
based on chi-square distribution is practically enough even when $n=50$.

\section{Detection of interactive loci pairs}
\label{sec:detection}

\subsection{Data analysis for the F${}_2$ population}
\label{subsec:f2}

As we explained in Section \ref{sec:intro},
\citet{Mizuta-etal10} conducted a genome scan of all pairs of marker loci of
$\mathrm{F}_2$ individuals of rice by using
chi-square statistics for independence.
In this section, we reexamine
the data from the viewpoint of multiple testings.

Rice has 12 chromosomes, and their total length is around 1600cM.
Two strains of rice used to produce the $\mathrm{F}_2$ population are 
Nipponbare and Kasalath.
Nipponbare is a short-grained rice in {\it japonica}\ variety, and
Kasalath is a long-grained rice in {\it indica}\ variety.
These two types have contrasting characteristics, and hence are used often in
QTL analysis.
By using Kasalath pollen, 
the $\mathrm{F}_1$ population was produced.
The $\mathrm{F}_2$ is an offspring resulting from the self-pollination of
$\mathrm{F}_1$ individuals.
The data comprise genotypes of 994
codominant markers at different locations covering the whole genome
for $n=186$ individuals of the $\mathrm{F}_2$ population
(\citet{Harushima-etal98}).

Figure \ref{fig:contour-f2} is a contour plot of chi-square statistics
calculated from all ${994 \choose 2}\doteq$ 500,000 marker pairs.
Because of linkage, the statistics are highly positively correlated,
and large values tend to appear in neighborhoods of
the ``high peak''.
 (As stated in Section \ref{sec:intro},
marker pairs on the same chromosome take large values.
Because these values simply measure the linkage, we ignore them.)

Table \ref{tab:largest-chi-squares} shows the highest 20 peaks
that do not to seem to be caused by the linkage effect.
The maximum chi-square statistic is 
\[
 \max_{1\le c_1<c_2\le 12} \max_{j_1,j_2} T_{c_1 c_2}(j_1,j_2) = 33.6
\]
observed between markers on chromosomes 9 and 12.
This corresponds to a $p$-value of $0.9\times 10^{-6}$
for a chi-square distribution with 4 degrees of freedom, which
is highly significant if we do not take the multiplicity of tests into account.
However, because of the high number of observed statistics (around 500,000),
some adjustment for multiplicity is required.
The Bonferroni-adjusted $p$-value for the maximum value is
$0.9\times 10^{-6}\times 500,000=0.45$.
However, this is conservative because the Bonferroni adjustment does not take
into account the highly positive correlations

\begin{table}[h]
\caption{The largest 20 chi-square values}
\label{tab:largest-chi-squares}
\begin{center}
\begin{small}
\begin{tabular}{r|lcr|lcr|c}
\thline
No.& Marker & Chr & (cM) & Marker & Chr &(cM) & Chi-square $T$ $^{1)}$ \\
\hline
 1 & R1683  & 9 &  94.1   & S10637A & 12 &  13.4 &   33.6 (2.9) \\
 2 & P130   & 6 &  54.0   & S12886  & 11 & 116.1 &   33.2 (7.1) \\
 3 & V163   & 5 &  71.1   & S11447  & 12 &  95.9 &   26.2 (1.2) \\
 4 & S2074  & 9 &  57.4   & S10906  & 10 &   2.0 &   23.8 (7.2) \\
 5 & P60    & 3 &  92.1   & S2572   & 12 &  26.5 &   23.3 (3.1) \\
 6 & Y5714L & 1 &  69.1   & R3203   &  1 & 160.0 &   21.7 (3.9) \\
 7 & S1046  & 1 & 161.9   & C946    &  4 &  10.4 &   20.9 (2.9) \\
 8 & V10A   & 3 &   2.5   & V133    &  8 & 107.0 &   20.7 (6.1) \\
 9 & C191A  & 1 & 141.9   & C1219   &  3 & 157.1 &   20.6 (1.7) \\
10 & P61    & 1 & 181.7   & R2965   & 10 &   2.3 &   20.5 (5.9) \\
11 & S11214 & 1 &  45.6   & S1520   &  6 &  15.2 &   20.0 (21.1) \\
12 & G55    & 3 &  34.4   & P126    &  6 &  39.6 &   19.8 (7.6) \\
13 & S1046  & 1 & 161.9   & G267    &  4 & 111.2 &   19.8 (4.3) \\
14 & R3192  & 1 &  26.9   & C922A   &  1 & 121.0 &   19.7 (3.0) \\
15 & R19    & 3 &  98.2   & G7004   &  4 &  72.3 &   19.5 (9.3) \\
16 & P60    & 3 &  92.1   & C1424   &  6 & 112.1 &   19.3 (3.8) \\
17 & R2625  & 1 & 155.3   & S851    &  3 & 150.1 &   19.2 (2.3) \\
18 & C506   & 9 &  93.0   & Y1053R  & 10 &  34.6 &   19.1 (3.8) \\
19 & S10879 & 9 &  94.4   & C496    & 11 &  30.3 &   19.0 (2.8) \\
20 & C2523S & 7 &   8.8   & S2545   & 12 &  72.5 &   19.0 (1.7) \\
\thline
\end{tabular}
\\
$^{1)}$ Figures in parentheses are chi-square $T$'s in the second experiment.
\end{small}
\end{center}
\end{table}

When we consider a particular chromosome pair, say $(c_1,c_2)$,
the statistics $T_{c_1 c_2}(j_1,j_2)$
 ($j_1=1,\ldots,n_{c_1}$, $j_2=1,\ldots,n_{c_2}$) have the
correlation structure described in Proposition \ref{prop:chi-square-field} (a).
Hence, the asymptotic null distribution of the maximum for pairs
on the chromosome pair $(c_1,c_2)$ can be evaluated.
Furthermore, noting Proposition \ref{prop:chi-square-field} (b), which states
that statistics on the different pairs of chromosomes are asymptotically
independent, we can evaluate the multiplicity-adjusted $p$-values
for the maximum statistics over whole chromosomes as follows:
\begin{equation}
\label{pvalue}
 \mbox{
 $p$-value} =
 F\Bigl(\max_{1\le c_1<c_2\le 12} \max_{j_1,j_2} T_{c_1 c_2}(j_1,j_2) \Bigr),
\end{equation}
\[
 F(x) = 1 - \prod_{1\le c_1<c_2\le 12} \Bigl\{ 1 -
 P\Bigl(\max_{t_1\in T_{c_1},t_2\in T_{c_2}} Y(t_1,t_2)^2 \ge x \Bigr)\Bigr\},
\]
where $Y$ is a chi random field defined in (\ref{Y})
with $p=2$, $m=4$, and $\rho_{ki}$ in (\ref{rhok12}).
The locations (M) of markers on chromosome $i$ are denoted by
$T_i=\{d_{i0},\ldots,d_{i n_i}\}$.

The multiplicity-adjusted $p$-value (\ref{pvalue}) 
for the maximum chi-square of 33.6 was estimated as
0.068 (Monte Carlo), 
0.104 (renewal theory), 
and
0.240 (tube method).
In applying Theorem \ref{thm:renewal}, we substituted the average of
the marker spacing on chromosome $i$ for $D_i$.
All of the peaks listed in Table \ref{tab:largest-chi-squares}
were not significant at 5\%.

In the Monte Carlo method, random variables were generated
from the recurrence relations in (\ref{AR}). 
Computational time was 14 days and 8 hours for 10,000 iterations
using a supercomputer SGI Altix3700 and the R language.

\begin{rem}
\label{rem:permutation}
In QTL analysis, permutation tests are commonly used for
estimating the null distribution of the maximum LOD scores
 (\citet{Churchill-Doerge94}).
For our problem, we can propose the procedure described below:
The data set of the genotypes of all individuals is denoted by $\cal D$.
Let $\Pi$ be the set of all permutations of individual numbers.
Repeat steps (i)--(ii).
\begin{itemize}
\item[(i)]
Choose a permutation $\pi$ from $\Pi$ at random.
Let ${\cal D}_\pi$ be the data set $\cal D$ with their individual numbers
relabeled by the permutation $\pi$.
\item[(ii)]
Make cross-classified tables
between all markers of $\cal D$ and all markers of ${\cal D}_\pi$
by their genotypes
 (i.e., in Table \ref{tab:crosstable}, locus 1 is taken from $\cal D$,
 and locus 2 is taken from ${\cal D}_\pi$), 
calculate the chi-square statistics from the tables,
and find their maximum.
\end{itemize}
The null distribution of the maximum chi-square statistics can be
estimated as the empirical distribution of the maxima obtained in (ii).
\end{rem}
However, the method referred to in Remark \ref{rem:permutation} requires
at least as much computational time as that required for Monte Carlo.

Moreover, \citet{Mizuta-etal10} performed additional genome scan searches for
another $\mathrm{F}_2$ population of a similar sample size.
The chi-square statistics corresponding to the peaks detected in the initial experiment are listed
in the last column of Table \ref{tab:largest-chi-squares}.
Except for peak No.\ 11, all other peaks in Table \ref{tab:largest-chi-squares}
showed low values of the chi-square statistics in the second scan.

\subsection{Data analysis for the BC population}
\label{subsec:bc}

Furthermore,
\citet{Mizuta-etal10}
carried out an additional experiment
using the reciprocal BC population to Nipponbare.
This experiment can distinguish where the interaction occurs,
i.e., male gametophyte, female gametophyte, or zygote.
They selected 159 markers including those exhibiting large chi-square values
in the $\mathrm{F}_2$ data analysis, and examined
the genotypes of all pairs of these selected markers in the BC populations.

Compared with the $\mathrm{F}_2$, the types of BDM pairs that can be
detected from the BC population are limited.
On the other hand, the detection power (the power function of test)
for detectable pairs is expected to be higher.

The BC population is the experimental crossing population produced by
crossing strain A with the $\mathrm{F}_1$ made from strains A and B.
Note that there is some arbitrariness about
whether the $\mathrm{F}_1$ is used as 
the maternal parent or pollen parent.
The set of two BC populations corresponding to these two cases is called
the reciprocal BC.
Only genotype AB is observed in the $\mathrm{F}_1$ population.
Two types of genotypes, AA and AB, are observed in the BC population.
We abbreviate these two genotypes to A and H, respectively. 
The genotypes of two loci 1 and 2 are cross-classified as shown in
Table \ref{tab:crosstable-bc}.
The chi-square statistic for independence obtained from this table
has an asymptotic chi-square distribution with 1 degree of freedom
under the null hypothesis that there exists no BDM pair.

\begin{table}[h]
\caption{Cross table of genotypes in two loci (BC) \newline
 (The table attaining at the maximum chi-square is shown in parentheses)
}
\label{tab:crosstable-bc}
\begin{center}
\begin{tabular}{c|ccc}
\thline
locus 1 (Chr 6\ \ S1520) $\backslash$ locus 2 (Chr 1\ \ S11214) & A & H \\
\hline
A (Nipponbare) & $n_\mathrm{AA} (75) $ & $n_\mathrm{AH}$ (13) \\
H & $n_\mathrm{HA}$ (64) & $n_\mathrm{HH}$ (83) \\
\thline
\end{tabular}
\end{center}
\end{table}

The $2\times 2$ table showing the maximum value of the chi-square statistics
is given in Table 
\ref{tab:crosstable-bc} (in parentheses).
The maximum value is 39.6, which was observed between chromosomes 1 and 6
in the BC population with the $\mathrm{F}_1$ pollen parent.
The sample size was $n=235$.
This is the loci pair listed as No.\,11 in Table \ref{tab:largest-chi-squares}.
In another BC population with the $\mathrm{F}_1$ maternal parent,
no significant peak was observed.

In order to obtain the multiplicity-adjusted $p$-value for this maximum value,
we need the joint distribution of the chi-square statistics.
In the BC case,
we can prove a proposition similar to Proposition \ref{prop:chi-square-field}:
Part (a) of Proposition \ref{prop:chi-square-field} holds if
convergence in law (\ref{convergence}) is replaced with the convergence 
\[
 T_{12}(j_1,i_2) \Rightarrow Z_1(j_1,j_2)^2 \quad (n\to\infty).
\]
Part (b) of Proposition \ref{prop:chi-square-field} holds as it is.

The multiplicity-adjusted $p$-value is
$2.86\times 10^{-6}$ (renewal theory) 
and
$1.57\times 10^{-5}$ (tube method).
In either case, it is highly significant.
This suggests that this pair is a candidate of the BDM pair
that we are seeking for
and that the selection occurred in male gametophyte, pollen.
Actually, \citet{Mizuta-etal10} confirmed that
the male gametophyte selection of the unbearable genotype combination of
the true BDM pair occurred through failure of pollen germination,
and the reciprocal disruption of duplicated genes
in the two strains caused the BDM incompatibility.
Note that no other significant peaks were detected.

Finally, we discuss
why the interaction was not detected in the $\mathrm{F}_2$ but was in the BC.
As explained in 
Section \ref{subsec:proof-corr} 
 (see Lemma \ref{lm:decomposition} and succeeding descriptions),
the chi-square statistic with 4 degrees of freedom obtained from
Table \ref{tab:crosstable} can be asymptotically decomposed into
four chi-square components each with 1 degree of freedom.
One of the four components corresponds to the chi-square statistic
obtained from Table \ref{tab:crosstable-bc}.
However, in producing the BC population,
there is some arbitrariness about whether $\mathrm{F}_1$ is used as
mother or father,
and both cases are assumed to be included in the $\mathrm{F}_2$ population
each with a probability 1/2.
Since the sample sizes for the $\mathrm{F}_2$ and BC data were similar
 (around 200), if there was no other significant component except for
the one component with 1 degree of freedom detected in
Table \ref{tab:crosstable-bc} (in parentheses).
 it is convincing that
the chi-square statistic of 20.0 (Table \ref{tab:largest-chi-squares}, No.\,11) %
in the $\mathrm{F}_2$ is almost half of that of 39.6 in the BC population
(pollen parent is $\mathrm{F}_1$).
In conclusion, although the chi-square statistic with 4 degrees of freedom obtained
from $\mathrm{F}_2$ has statistical power in many directions,
larger sample size was needed to detect the BDM pair.

\section{Proofs}
\label{sec:proofs}

\subsection{Proof of Proposition \ref{prop:chi-square-field}}
\label{subsec:proof-corr}

First, we provide asymptotic presentations of chi-square statistics
for independence when the independent model is true.
Let $X=(x_{ij})_{a\times b}$ ($x_{\cdot\cdot}=n$) be a contingency table
distributed as a multinomial distribution with
the cell probability $(p_{ij})_{a\times b}$ $(p_{\cdot\cdot}=1)$.
Here, we apply the convention that the summation with respect to
an index is denoted by ``$\cdot$''.
The chi-square statistic for the hypothesis of independence
$H_0: p_{ij}=p_{i\cdot}{p_{\cdot j}}$ is denoted by
\[
 T = T(X) = \sum_{i,j}
   \frac{(x_{ij}-x_{i\cdot}x_{\cdot j}/n)^2}
        {x_{i\cdot}x_{\cdot j}/n}.
\]
The proofs of the following lemmas are easy and omitted.

\begin{lm}
\label{lm:decomposition}
For a $3\times 3$ table $X=(x_{ij})_{1\le i,j\le 3}$, define four $2\times 2$ tables:
\[
X_1 = \begin{pmatrix} x_{11} & x_{12} \\
                      x_{21} & x_{22} \end{pmatrix}, \quad
X_2 = \begin{pmatrix} x_{11}+x_{12} & x_{13} \\
                      x_{21}+x_{22} & x_{23} \end{pmatrix}, \quad
X_3 = \begin{pmatrix} x_{11}+x_{21} & x_{12}+x_{22} \\
                      x_{31}        & x_{32} \end{pmatrix},
\]
\[
X_4 = \begin{pmatrix} x_{11}+x_{12}+x_{21}+x_{22} & x_{13}+x_{23} \\
                      x_{31}+x_{32}               & x_{33} \end{pmatrix}.
\]
Under $H_0$, four statistics
$T(X_1), T(X_2), T(X_3), T(X_4)$ are asymptotically distributed
according to the independent chi-square distributions
with $1$ degree of freedom, and
it holds that
\[
 T(X) = T(X_1) + T(X_2) + T(X_3) + T(X_4) + O_p(n^{-1/2}).
\]
\end{lm}

\begin{lm}
For a $2\times 2$ table $X=(x_{ij})_{1\le i,j\le 2}$ with the cell probability $(p_{ij})_{1\le i,j\le 2}$,
\begin{equation}
T(X)
 = \frac{1}{n}
 \Biggl( \sum_{i,j=1}^2 (-1)^{i+j} \sqrt{
  \frac{p_{3-i,\cdot}p_{\cdot,3-j}}{p_{i\cdot}p_{\cdot j}}} \, x_{ij}
 \Biggr)^2 +O_p(n^{-1/2})
\label{T'}
\end{equation}
holds under $H_0$.
\end{lm}

For the $\mathrm{F}_2$ individuals $t=1,\ldots,n$ made from two strains A and B,
by cross-classifying the genotypes of
marker $i$ ($i=1,\ldots,m$) on chromosome 1 and
marker $j$ ($j=1,\ldots,\widetilde m$) on chromosome 2,
we have the $3\times 3$ tables represented by Table \ref{tab:crosstable}.
Let $T_{ij}$ be the chi-square statistic obtained from the table
for marker pair $(i,j)$.

For individual $t$,
let $\epsilon_i^{(t)}$ be the genotype of locus $i$ on
chromosome 1 inherited from its mother, and
let $\delta_i^{(t)}$ be that from its father.
Let $\widetilde\epsilon_j^{(t)}$ be the genotype of locus $j$ on
chromosome 2 inherited from its mother, and
let $\widetilde\delta_j^{(t)}$ be that from its father.
We let
\[
 \epsilon_i^{(t)},\,\delta_i^{(t)}\,
   \widetilde\epsilon_j^{(t)},\,\widetilde\delta_j^{(t)}\,
 = \begin{cases} 1 & (\mbox{from strain A}), \\
                -1 & (\mbox{from strain B}). \end{cases}
\] 
Then, the $4n$ random vectors
$\bigl(\epsilon_1^{(t)},\ldots,\epsilon_m^{(t)}\bigr)$,
$\bigl(\delta_1^{(t)},\ldots,\delta_m^{(t)}\bigr)$,
$\bigl(\widetilde\epsilon_1^{(t)},\ldots,
       \widetilde\epsilon_{\widetilde m}^{(t)}\bigr)$,
$\bigl(\widetilde\delta_1^{(t)},\ldots,
       \widetilde\delta_{\widetilde m}^{(t)}\bigr)$, $t=1,\ldots,n$
are independent of each other, and all elements take the value $\pm 1$
with probabilities 1/2 and 1/2 
satisfying a Markov property
\begin{align}
\label{markov}
 P\bigl(\epsilon_{i+1}^{(t)}=\pm\epsilon_i^{(t)}\,\big|\,\epsilon_i^{(t)}\bigr)
 =
 P\bigl(\delta_{i+1}^{(t)}=\pm\delta_i^{(t)}\,\big|\,\delta_i^{(t)}\bigr)
 =
 \frac{1}{2}\bigl( 1\pm e^{-2 d_{i,i+1}} \bigr).
\end{align}
$\widetilde\epsilon_j^{(t)}$ and $\widetilde\delta_j^{(t)}$ have the same
Markov structure with $d_{i,i+1}$ replaced by $\widetilde d_{j,j+1}$.
Here, the genetic distance between markers $i$ and $i'$ on chromosome 1 is
denoted by $d_{ii'}$ (M),
and the genetic distance between markers $j$ and $j'$ on chromosome 2 is
denoted by $\widetilde d_{jj'}$ (M).
This assumption of linkage is called Haldane's model. 
From this model, it is easy to derive the correlation structures
\[
   E\bigl[\epsilon_i^{(t)}\epsilon_{i'}^{(t)}\bigr]
 = E\bigl[\delta_i^{(t)}\delta_{i'}^{(t)}\bigr] = e^{-2 d_{ii'}}, \qquad
   E\bigl[\widetilde\epsilon_j^{(t)}\widetilde\epsilon_{j'}^{(t)}\bigr]
 = E\bigl[\widetilde\delta_j^{(t)}\widetilde\delta_{j'}^{(t)}\bigr]
 = e^{-2 \widetilde d_{jj'}}.
\] 

Using this notation, the $3\times 3$ table represented by Table \ref{tab:crosstable}
can be rewritten as
\begin{align}
&
\begin{pmatrix}
 n_\mathrm{AA} & n_\mathrm{AB} & n_\mathrm{AH} \\
 n_\mathrm{BA} & n_\mathrm{BB} & n_\mathrm{BH} \\
 n_\mathrm{HA} & n_\mathrm{HB} & n_\mathrm{HH}
\end{pmatrix} 
 = \sum_{t=1}^n
\begin{pmatrix}
 \frac{1}{4}(1+\epsilon_i^{(t)})(1+\delta_i^{(t)}) \\
 \frac{1}{4}(1-\epsilon_i^{(t)})(1-\delta_i^{(t)}) \\
 \frac{1}{2}(1-\epsilon_i^{(t)}\delta_i^{(t)})
\end{pmatrix}
\nonumber \\ 
& \qquad\qquad \times 
\begin{pmatrix}
 \frac{1}{4}(1+\widetilde\epsilon_j^{(t)})(1+\widetilde\delta_j^{(t)}) &
 \frac{1}{4}(1-\widetilde\epsilon_j^{(t)})(1-\widetilde\delta_j^{(t)}) &
 \frac{1}{2}(1-\widetilde\epsilon_i^{(t)}\widetilde\delta_i^{(t)})
\end{pmatrix}.
\label{3x3}
\end{align}
In order to derive the joint distribution of the chi-square statistics $T_{ij}$,
we decompose the $3\times 3$ table into four $2\times 2$ tables (i)--(iv)
according to Lemma \ref{lm:decomposition}.

(i) Table
$
\begin{pmatrix}
 n_\mathrm{AA} & n_\mathrm{AB} \\
 n_\mathrm{BA} & n_\mathrm{BB}
\end{pmatrix}.
$
The sum of the expected frequencies is $n/4$.
From (\ref{T'}),
the corresponding chi-square statistic has the asymptotic representation
\begin{align*}
T_{1,ij} 
&= \frac{1}{n/4}(n_\mathrm{AA} - n_\mathrm{AB}
 - n_\mathrm{BA} + n_\mathrm{BB})^2 + O_p(n^{-1/2}) \\
&= \left( \frac{1}{\sqrt{n}}\sum_{t=1}^n z_{1,ij}^{(t)} \right)^2
    + O_p(n^{-1/2}), \quad
 z_{1,ij}^{(t)} =
   (\epsilon_i^{(t)}+\delta_i^{(t)})
   (\widetilde\epsilon_j^{(t)}+\widetilde\delta_j^{(t)})/2.
\end{align*}

(ii) Table
$
\begin{pmatrix}
 n_\mathrm{AA} + n_\mathrm{AB} & n_\mathrm{AH} \\
 n_\mathrm{BA} + n_\mathrm{BB} & n_\mathrm{BH}
\end{pmatrix}.
$
The sum of the expected frequencies is $n/2$.
The corresponding chi-square statistic has the asymptotic representation
\begin{align*}
T_{2,ij}
& = \frac{1}{n/2}\left((n_\mathrm{AA} + n_\mathrm{AB}) - n_\mathrm{AH}
 -(n_\mathrm{BA} + n_\mathrm{BB}) + n_\mathrm{BH}\right)^2
 + O_p(n^{-1/2}) \\
& = \left( \frac{1}{\sqrt{n}}\sum_{t=1}^n z_{2,ij}^{(t)} \right)^2
  + O_p(n^{-1/2}), \quad
 z_{2,ij}^{(t)} =
   (\epsilon_i^{(t)}+\delta_i^{(t)})
   (\widetilde\epsilon_j^{(t)}\widetilde\delta_j^{(t)})/\sqrt{2}.
\end{align*}

(iii) Table
$
\begin{pmatrix}
 n_\mathrm{AA}+n_\mathrm{BA} & n_\mathrm{AB}+n_\mathrm{BB} \\
 n_\mathrm{HA}               & n_\mathrm{HB}
\end{pmatrix}.
$
The sum of the expected frequencies is $n/2$.
The corresponding chi-square statistic has the asymptotic representation
\begin{align*}
T_{3,ij}
& = \frac{1}{n/2}\left(
 (n_\mathrm{AA}+n_\mathrm{BA}) - (n_\mathrm{AB}+n_\mathrm{BB})
 - n_\mathrm{HA} + n_\mathrm{HB}\right)^2 + O_p(n^{-1/2}) \\
& = \left( \frac{1}{\sqrt{n}}\sum_{t=1}^n z_{3,ij}^{(t)} \right)^2
 + O_p(n^{-1/2}), \quad
 z_{3,ij}^{(t)} =
   (\epsilon_i^{(t)}\delta_i^{(t)})
   (\widetilde\epsilon_j^{(t)}+\widetilde\delta_j^{(t)})/\sqrt{2}.
\end{align*}

(iv) Table
$
\begin{pmatrix}
 n_\mathrm{AA}+n_\mathrm{AB}+n_\mathrm{BA}+n_\mathrm{BB}
                             & n_\mathrm{AH}+n_\mathrm{BH} \\
 n_\mathrm{HA}+n_\mathrm{HB} & n_\mathrm{HH}
\end{pmatrix}.
$
The sum of the expected frequencies is $n$.
The corresponding chi-square statistic has the asymptotic representation
\begin{align*}
T_{4,ij}
& = \frac{1}{n}\left((n_\mathrm{AA}+n_\mathrm{AB}+n_\mathrm{BA}+n_\mathrm{BB})
 - (n_\mathrm{AH}+n_\mathrm{BH}) - (n_\mathrm{HA}+n_\mathrm{HB})
 + n_\mathrm{HH}\right)^2 \\ & \quad + O_p(n^{-1/2}) \\
& = \left( \frac{1}{\sqrt{n}}\sum_{t=1}^n z_{4,ij}^{(t)} \right)^2
 + O_p(n^{-1/2}), \quad
   z_{4,ij}^{(t)} = \epsilon_i^{(t)}\delta_i^{(t)}
                    \widetilde\epsilon_j^{(t)}\widetilde\delta_j^{(t)}.
\end{align*}

$z_{k,ij}^{(t)}$ ($k=1,2,3,4$) has a mean 0 and a covariance structure
\begin{align*}
E\bigl[z_{1,ij}^{(t)} z_{1,i'j'}^{(t)}\bigr]
&= 
E\bigl[(\epsilon_i^{(t)}+\delta_i^{(t)})
       (\epsilon_{i'}^{(t)}+\delta_{i'}^{(t)})\bigr]
E\bigl[(\widetilde\epsilon_j^{(t)}+\widetilde\delta_j^{(t)})
  (\widetilde\epsilon_{j'}^{(t)}+\widetilde\delta_{j'}^{(t)})\bigr]/4 \\
&= e^{-2 d_{ii'}} e^{-2 \widetilde d_{jj'}}, \\
E\bigl[z_{2,ij}^{(t)} z_{2,i'j'}^{(t)}\bigr]
&=
E\bigl[(\epsilon_i^{(t)}+\delta_i^{(t)})
       (\epsilon_{i'}^{(t)}+\delta_{i'}^{(t)})\bigr]
E\bigl[(\widetilde\epsilon_j^{(t)}\widetilde\delta_j^{(t)})
  (\widetilde\epsilon_{j'}^{(t)}\widetilde\delta_{j'}^{(t)})\bigr]/2 \\
&= e^{-2 d_{ii'}} e^{-4 \widetilde d_{jj'}}, \\
E\bigl[z_{3,ij}^{(t)} z_{3,i'j'}^{(t)}\bigr]
&=
E\bigl[(\epsilon_i^{(t)}\delta_i^{(t)})
       (\epsilon_{i'}^{(t)}\delta_{i'}^{(t)})\bigr]
E\bigl[(\widetilde\epsilon_j^{(t)}+\widetilde\delta_j^{(t)})
  (\widetilde\epsilon_{j'}^{(t)}+\widetilde\delta_{j'}^{(t)})\bigr]/2 \\
&= e^{-4 d_{ii'}} e^{-2 \widetilde d_{jj'}}, \\
E\bigl[z_{4,ij}^{(t)} z_{4,i'j'}^{(t)}\bigr]
&=
E\bigl[(\epsilon_i^{(t)}\delta_i^{(t)})
       (\epsilon_{i'}^{(t)}\delta_{i'}^{(t)})\bigr]
E\bigl[(\widetilde\epsilon_j^{(t)}\widetilde\delta_j^{(t)})
  (\widetilde\epsilon_{j'}^{(t)}\widetilde\delta_{j'}^{(t)})\bigr]
 = e^{-4 d_{ii'}} e^{-4 \widetilde d_{jj'}}, \\
E\bigl[z_{k,ij}^{(t)} z_{k',i'j'}^{(t)}\bigr]
&= 0 \quad (k\ne k').
\end{align*}
Part (a) of Proposition \ref{prop:chi-square-field} follows from
the central limit theorem and the continuous mapping theorem.

When
markers $i$ and $i'$ are on different chromosomes, or
markers $j$ and $j'$ are on different chromosomes,
we can let $d_{ii'}=\infty$ or $\widetilde d_{jj'}=\infty$.
In each case, $E\bigl[z_{k,ij}^{(t)} z_{k',i'j'}^{(t)}\bigr]=0$
for all $k$ and $k'$.
This implies that the statistics $T_{ij}$ and $T_{i'j'}$ are
made from random variables whose limiting distributions are independent Gaussian,
and hence, part (b) of Proposition \ref{prop:chi-square-field} follows.

\subsection{Proof of Theorem \ref{thm:renewal}}
\label{subsec:proof-renewal}

The proof is divided into three parts.
Section \ref{subsubsec:renewal} provides an outline of the proof
without proving a key relation (\ref{tmp}).
In Section \ref{subsubsec:tmp},
it is shown that the chi field $Y(t)$ restricted on lattice points is
approximated by a suitably defined random walk,
and that the maximum of $Y(t)$ can be approximated by the maximum of
the corresponding random walk ((\ref{eq3}) and (\ref{M})).
Then, (\ref{tmp}) is proved using an identity of Laplace transform
provided in Section \ref{subsubsec:extension}.
Differently from changepoint problems dealt with in previous work,
the random field $Y(t)$ has a general dimensional index set and general
degrees of freedom.
We thereby need to introduce a random walk on a general dimensional
index set,
and an integral on a general dimensional unit sphere.

\subsubsection{Proof of (\ref{renewal})}
\label{subsubsec:renewal}

By arranging the index set $J$ in the lexicographic order,
we can let $j^0=(j_1^0,\ldots,j_d^0)\in J$ be the first point such that
the random field $Y(jD)$ takes a value of at least $b$.
Let
\begin{align*}
J^0(j^0) = \bigl\{ j \in J \,|\,
&   j_1>j_1^0, \\
& \mbox{or}\ \ j_1=j_1^0,\ j_2>j_2^0, \\
& \mbox{or}\ \ \ldots, \\
& \mbox{or}\ \ j_1=j_1^0,\ \ldots,\ j_{d-1}=j_{d-1}^0,\ j_d>j_d^0 \bigr\}.
\end{align*}

Let $\S^{m-1}$ be the unit sphere in $\R^m$.
Let $du$ be its volume element at $u\in\S^{m-1}$.
Let $dy=(y,y+dy)$.

The event $\bigl\{\max_{j\in J} Y(j D) \ge b\bigr\}$ is exclusively divided
by the value of $j^0\in J$
(see, e.g.,
\citet{Dupuis-Siegmund00}, (15)) as
\begin{align}
P\Bigl( & \max_{j\in J} Y(j D) \ge b \Bigr) \nonumber \\
& =\ %
 \sum_{j^0\in J}
 P\biggl( \max_{j\in J^0(j^0)} Y(j D) < b,\ Y(j^0 D) \ge b \biggr) \nonumber \\
& =\ %
 \int_{\S^{m-1}} \sum_{j^0\in J}
 P\biggl( \max_{j\in J^0(j^0)} Y(j D) < b,\ Y(j^0 D) \ge b,\ %
 \frac{Z(j^0 D)}{Y(j^0 D)} \in du \biggr) \nonumber \\
& = %
 \int_{y>b} \, \int_{\S^{m-1}} \sum_{j^0\in J}
 P\biggl( \max_{j\in J^0(j^0)} Y(j D) < b,\ Y(j^0 D) \in dy,\ %
 \frac{Z(j^0 D)}{Y(j^0 D)} \in du \biggr) \nonumber \\
& =\ %
 \int_{y>b} \,
 \int_{\S^{m-1}} \sum_{j^0\in J}
 P\biggl( \max_{j\in J^0(j^0)} Y(j D) < b\mid Z(j^0 D)=y u \biggr) \nonumber \\
& \qquad\qquad\qquad\times
 P\biggl( Y(j^0 D) \in dy,\ \frac{Z(j^0 D)}{Y(j^0 D)}
 \in du \biggr) \nonumber \\
& =\ %
 \int_{x>0} \,
 \int_{\S^{m-1}} \sum_{j^0\in J}
 P\biggl( \max_{j\in J^0(j^0)} Y(j D) < b\mid Z(j^0 D)=y u \biggr) \nonumber \\
& \qquad\qquad\qquad\times
 P\biggl( Y(j^0 D) \in \Bigl(b+\frac{(x,x+dx)}{b} \Bigr),\ %
 \frac{Z(j^0 D)}{Y(j^0 D)} \in du \biggr).
\label{tmp1}
\end{align}
In the last expression, we made change of variable $y=b+x/b$.

For fixed $j^0$, $Z_k(j^0 D)\sim N_m(0,I_m)$, and hence
$Y(j^0 D)\sim\chi_m$ and \\ $Z(j^0 D)/Y(j^0 D)\sim\Unif(\S^{m-1})$ are
independent.  Therefore,
\begin{align}
 P\biggl( Y( & j^0 D) \in \Bigl(b+\frac{(x,x+dx)}{b} \Bigr),\ %
 \frac{Z(j^0 D)}{Y(j^0 D)} \in du \biggr) \nonumber \\
& =\,%
 P\biggl( Y(j^0 D)^2 \in \Bigl((b+x/b)^2, (b+x/b)^2 \cdot 2 dx \Bigr)
 \biggr) \times \frac{du}{\Vol(\S^{m-1})} \nonumber \\
& =\,%
 \frac{2}{2^{m/2}\Gamma(m/2)} b^{m-2} e^{-b^2/2} e^{-x} dx
 \times \frac{du}{\Vol(\S^{m-1})}.
\label{tmp2}
\end{align}

Moreover, as shown later,
\begin{equation}
\int_{x>0}  
 P\biggl( \max_{j\in J^0(j^0)} Y(j D) < b\mid Z(j^0 D)=y u \biggr) dx
 \,\sim\, \prod_i \bar\rho_i c_i^2 \nu(c_i\sqrt{2\bar\rho_i})
\label{tmp}
\end{equation}
($y=b+x/b$, $\bar\rho_i=\bar\rho_i(u)$ is in (\ref{rhobar})).

By substituting (\ref{tmp2}) and (\ref{tmp}) into (\ref{tmp1}) and
noting that
$\prod_i D_i \sum_{j^0\in J}\sim
 \int_{\widetilde T} \prod_i dt_i =|\widetilde T|$,
$\Vol(\S^{m-1})=2\pi^{m/2}/\Gamma(m/2)$,
we obtain
\begin{align*}
P\Bigl( \max_{j\in J} & Y(j D) \ge b \Bigr) \ %
\\ &
\sim\ \frac{|\widetilde T|}{\prod_i D_i} \times
\frac{1}{(2\pi)^{m/2}} b^{m-2} e^{-b^2/2} \int_{\S^{m-1}} du
 \prod_i \bar\rho_i c_i^2 \nu(c_i\sqrt{2\bar\rho_i}).
\end{align*}
This means (\ref{renewal}).

\subsubsection{Proof of (\ref{tmp})}
\label{subsubsec:tmp}

We use the large-deviation approach developed by \citet{Siegmund88}.
See also \citet{Kim-Siegmund89}.

Suppose that $t$ is fixed.
Under a conditional probability measure given
$Z(t) = (Z_k(t))_{1\le k\le m} = \xi = (\xi_k)_{1\le k\le m}$,
the $\R^m$-valued random field $Z(t+h)=(Z_k(t+h))_{1\le k\le m}$
with the index $h=(h_i)_{1\le i\le p}$
is a Gaussian random field with a mean of
\[
 E[ Z_k(t+h) \,|\, \xi] = R_k(h) \xi_k,
\]
and a covariance function of
\[
 \Cov( Z_k(t+h), Z_{k'}(t+h') \,|\, \xi) =
\begin{cases}
 R_k(h-h') - R_k(h) R_k(h') & (k=k'), \\
 0                          & (k\ne k').
\end{cases}
\]
When $h_i$ is small, these moments can be rewritten as
\begin{align*}
E[ Z_k(t+h) \,|\, \xi]
& = \xi_k - \xi_k \sum_{i=1}^p \rho_{ki}|h_i| + \xi_k o(|h|), \\
\Cov(Z_k(t+h), Z_k(t+h') \,|\,\xi)
& = \sum_{i=1}^p \rho_{ki}(|h_i|+|h'_i|-|h_i-h'_i|) + o(|h|).
\end{align*}
We consider asymptotics where
\[
 h_i\to 0,\ \ \Vert\xi\Vert\to\infty \quad\mbox{such that}\quad \xi_k/\Vert\xi\Vert = u_k,\ \ \Vert\xi\Vert \sqrt{h_i} = O(1).
\]
Since
$ Z_k(t+h) = \xi_k  + O(\sqrt{|h|}) = \xi_k (1 + O(|h|)) $,
we have
\begin{align*}
Y(t+h)
&= \sqrt{\sum_{k=1}^m Z_k(t+h)^2} \\
&= \Vert \xi\Vert
 \sqrt{1 + \frac{\sum_k (Z_k(t+h)^2 - \xi_k^2)}{\Vert \xi\Vert^2}} \\
&= \Vert \xi\Vert \biggl\{
 1 + \sum_k \frac{\xi_k (Z_k(t+h) - \xi_k)}{\Vert \xi\Vert^2}
 (1 + O(|h|)) + O(|h|^2) \biggr\} \\
&= \Vert \xi\Vert
 + \frac{1}{\Vert\xi\Vert}\sum_k \xi_k (Z_k(t+h) - \xi_k)
 (1 + O(|h|)).
\end{align*}
In this expression, we used
\[
 Z_k(t+h)^2 - \xi_k^2 = 2\xi_k (Z_k(t+h) - \xi_k) (1 + O(|h|)) = O(1)
\]
and $ \xi_k (Z_k(t+h) - \xi_k)/\Vert \xi\Vert^2 = O(|h|) $.
Next, consider a conditional random field with the index $h$ defined by 
$\Vert\xi\Vert \bigl\{ Y(t+h) - \Vert\xi\Vert \bigr\} \Big|_{Z(t)=\xi}$.
The leading terms of the mean and covariance function of this field are shown to be
\begin{equation}
\label{mean-cov}
 - \sum_k \Vert\xi\Vert^2 u_k^2 \sum_i \rho_{ki}|h_i|, \qquad
 \sum_k \Vert\xi\Vert^2 u_k^2 \sum_i \rho_{ki}(|h_i|+|h'_i|-|h_i-h'_i|), 
\end{equation}
respectively.

From now on,
let $t=j^0 D$ and $h=(j-j^0) D$ in the multi-index notation of (\ref{multiindex}),
and consider the following (finite dimensional) joint distribution
under the condition that $Z(j^0 D)=\xi$:
\begin{equation}
\label{eq2}
 b \bigl\{ Y(j D) - \Vert\xi\Vert \bigr\} \Big|_{Z(j^0 D)=\xi},
 \quad j=(j_1,\ldots,j_p)\in J \subset \Z^p.
\end{equation}
When
\[
 \Vert\xi\Vert,b\to\infty, \quad D_i\to 0 \quad\mbox{such that}\ \ \Vert\xi\Vert\sim b, \quad b\sqrt{D_i}\to c_i\in (0,\infty),
\]
from (\ref{mean-cov}), the limit of the conditional mean is
\[
 -\sum_k u_k^2 \sum_i \rho_{ki} c_i^2 |j_i| =  - \sum_i \bar\rho_i c_i^2 |j_i|
\]
with $\bar\rho_i=\bar\rho_i(u)$ defined in (\ref{rhobar}),
and the limit of the covariance between
$b \bigl\{ Y(j D) - \Vert\xi\Vert \bigr\}$ and
$b \bigl\{ Y(j' D) - \Vert\xi\Vert \bigr\}$ ($j'=(j'_1,\ldots,j'_p)$) is
\begin{align*}
 \sum_k u_k^2 \sum_i \rho_{ki}c_i^2 & (|j_i|+|j'_i| -|j_i-j'_i|)
 = \sum_i \bar\rho_{ki}c_i^2(|j_i|+|j'_i|-|j_i-j'_i|) \\
&= \begin{cases} 2 \sum_i \bar\rho_{ki}c_i^2 \min(|j_i|,|j'_i|)
  & \mbox{($j_i$ and $j'_i$ have the same sign)}, \\ 0 & \mbox{(otherwise)}.
\end{cases}
\end{align*}

Since the limit becomes Gaussian again,
the limiting distribution of (\ref{eq2}) is equivalent to the distribution of
\[
 \sum_{i=1}^p (S^+_{i j_i} + S^-_{i j_i}), \quad j=(j_1,\ldots,j_p) \in J,
\]
where
\begin{align*}
& \displaystyle
 S^+_{i t} = \begin{cases}
 X_{i1} + \cdots + X_{it} & (t>0), \\
 0                        & \mbox{(otherwise)}, \end{cases} 
\\
& \displaystyle
 S^-_{i t} = \begin{cases}
 X_{i,-1} + \cdots + X_{i,t} & (t<0), \\
 0                            & \mbox{(otherwise)}, \end{cases}
\end{align*}
with
$X_{it} \sim N(-\bar\rho_i c_i^2, 2 \bar\rho_i c_i^2)$
 ($i=1,\ldots,p$, $t\in \Z$) being independent Gaussian random variables.

Summarizing the discussion above,
we have proved that for $y=\Vert\xi\Vert=b+x/b\sim b$,
\begin{align}
P\biggl( \max_{j\in J^0(j^0)} & Y(j D) < b \mid Z(j^0 D) = y u \biggr)
 \nonumber \\
 & =\,
 P\biggl( \max_{j\in J^0(j^0)}
 b \bigl\{ Y(j D) - \xi \bigr\} < -x \mid Z(j^0 D) = \xi \biggr) \nonumber \\
& \sim\,
 P\biggl( \max_{j\in J^0(j^0)} \sum_{i=1}^p S_{i,j_i} < -x \biggr).
\label{eq3}
\end{align}

In what follows, let $j:=j-j^0$ for simplicity.
$j\in J^0(j^0)$ is rewritten as $j\in J^0(0)$.
Let
\[
 M^+_i = \max_{j>0} S_{ij}, \qquad M^-_i = \max_{j\le 0} S_{ij}.
\]
Because of
\[
\max_{j\in J^0(0)}
= \max\biggl[
\max_{j_1>0,\,j_2,\ldots,j_p\in\Z},
\max_{j_1=0,\,j_2>0,\,j_3,\ldots,j_p\in\Z},
\ldots,
\max_{j_1=j_2=\cdots=j_{p-1}=0,\,j_p>0}\biggr],
\]
the event
\begin{equation}
\label{max}
 \max_{j\in J^0(0)} \sum_{i=1}^p S_{i,j_i} < -x
\end{equation}
is equivalent to the event that all of the following inequalities hold:
\begin{align*}
M^+_1 + \max\{M^+_2,M^-_2\} + \max\{M^+_3,M^-_3\} + \cdots +  
 \max\{M^+_p,M^-_p\} & < -x, \\
M^+_2 + \max\{M^+_3,M^-_3\} + \cdots + \max\{M^+_p,M^-_p\} & < -x, \\
   \ldots & \\
M^+_p & < -x.
\end{align*}

Since $M^-_p\ge 0$, if both
\[
M^+_i + \max\{M^+_{i+1},M^-_{i+1}\} + \cdots + \max\{M^+_{p-1},M^-_{p-1}\}
 + M^-_p < -x
\]
and
$ M^+_p < -x $
hold, then
\begin{align*}
 M^+_i + \max\{M^+_{i+1},M^-_{i+1}\} + \cdots + & \max\{M^+_{p-1},M^-_{p-1}\} + M^+_p
\\ 
& < -x - M^-_p + M^+_p \\
&
 < -2x < -x
\end{align*}
holds.  This implies that
\[
 M^+_i + \max\{M^+_{i+1},M^-_{i+1}\} + \cdots + \max\{M^+_{p-1},M^-_{p-1}\}
 + \max\{M^+_p,M^-_p\} < -x.
\]
Therefore,
(\ref{max}) is equivalent to the event that all of the following hold:
\begin{align*}
M^+_1 + \max\{M^+_2,M^-_2\} 
 + \cdots + \max\{M^+_{p-1},M^-_{p-1}\} + M^-_p & < -x, \\
M^+_2 
 + \cdots + \max\{M^+_{p-1},M^-_{p-1}\} + M^-_p & < -x, \\
   \ldots & \\
M^+_p & < -x.
\end{align*}
Repeating this argument reveals that (\ref{max}) is equivalent to
the event that all of the following inequalities hold:
\begin{align*}
M^+_1 + M^-_2 + M^-_3 + \cdots + M^-_p & < -x, \\
M^+_2 + M^-_3 + \cdots + M^-_p & < -x, \\
   \ldots & \\
M^+_p & < -x.
\end{align*}
That is,
\begin{align}
(\ref{eq3})
&\,\sim\,
 P\Bigl( M^+_i + M^-_{i+1} + \cdots + M^-_p < -x,\ 1\le i\le p \Bigr)
\nonumber \\
&\,=\,
 P\Bigl( \max_{1\le i\le p}
   \bigl( M^+_i + M^-_{i+1} + \cdots + M^-_p \bigr) < -x\Bigr).
\label{M}
\end{align}

Because the mean $\mu_i$ and variance $\sigma^2_i$ of $X_{ik}$ satisfy
\[
 \frac{-\mu_i}{\sigma^2_i} = \frac{-\bar\rho_i c_i^2}{2\bar\rho_i c_i^2} \equiv -\frac{1}{2}, 
\]
it follows for any $p\ge 1$ that
\begin{align}
 \int_0^\infty e^{-x}
 P\Bigl( & \max_{1\le i\le p} \bigl( M^+_i + M^-_{i+1} + \cdots + M^-_p \bigr) < -x \Bigr)\,dx
\nonumber \\
&= \prod_{i=1}^m \mu_i \nu(\mu_i/\sigma_i)
= \prod_{i=1}^m \rho_i c_i^2 \nu(c_i\sqrt{2\rho_i}).
\label{extension}
\end{align}
A proof is given below.
Combining (\ref{eq3}), (\ref{M}) and (\ref{extension}) yields (\ref{tmp}).

\subsubsection{Proof of (\ref{extension})}
\label{subsubsec:extension}

Note that $M^+_1,\,M^-_1,\,\ldots,\,M^+_p,\,M^-_p$ are all independent.
A proof of $p=1$ is given by \citet{Siegmund92}, Lemma 19. 
For $p\ge 2$, from the integration by parts essentially proved by
\citet{Siegmund92}, Proposition 24, 
we have
\begin{align*}
\mbox{RHS}&\mbox{ of (\ref{extension})} \\
&
= \int_0^\infty e^{-x}
 P\Bigl( \max_{1\le i\le p}\bigl( M^+_i + M^-_{i+1} + \cdots + M^-_p \bigr) < -x \Bigr)\,dx \\
&=
 \int_0^\infty e^{-x}
 P\Bigl( \max_{1\le i\le p-1}\bigl( M^+_i + M^-_{i+1} + \cdots + M^-_p \bigr) < -x \Bigr)
 P\Bigl( M^+_p < -x \Bigr)\,dx \\
&= \mu_p \nu(2\mu_p/\sigma_p)
 \int_0^\infty e^{-x}
 P\Bigl( \max_{1\le i\le p-1}\bigl( M^+_i + M^-_{i+1} + \cdots + M^-_{p-1} \bigr) < -x \Bigr)
 \,dx.
\end{align*}
The proof follows from mathematical induction.

\subsection{Proof of Theorem \ref{thm:tube}}
\label{subsec:proof-tube}

\subsubsection{Random fields defined by triangulation}

First, we discuss in detail the construction of $\widetilde Z_k$ 
by triangulation of index set. 
It is well known that a $p$-dimensional cube $[0,1]^p$ can be dissected
into congruent $p!$ simplices.
For example, let $\Pi_p$ be the set of all permutations of $\{1,\ldots,p\}$,
and for each $\pi\in\Pi_p$ let
\[
 S_\pi = \{ (x_1,\ldots,x_p) \in [0,1]^p \mid x_{\pi(1)}\ge \cdots \ge x_{\pi(p)} \}.
\]
Then, $[0,1]^p=\bigcup_{\pi\in\Pi_p} S_\pi$, and
$S_\pi$ and $S_{\pi'}$ ($\pi\ne\pi'$) do not share any interior point.

We dissect the $p$-dimensional rectangle
whose vertices are flanking lattice points 
\[
 [d_{1 j_1-1},d_{1 j_1}]\times\cdots\times[d_{p j_p-1},d_{p j_p}]
\]
into $p!$ simplices according to the same rule.
Let $e_i\in\R^p$ be a vector whose elements are all 0
except for the $i$th element of the value 1.
Write
\[
 t_0 = (t_{1 j_1-1},\ldots,t_{p j_p-1}), \quad D_i = D_{i j_i} = t_{i j_i}-t_{i j_i-1} \quad (i=1,\ldots,p)
\]
for simplicity.
Then, one of the resulting simplices produced by the dissection is
\begin{equation}
\label{simplex}
 \conv\Bigl\{t_0+\sum_{l=1}^i D_l e_l \mid i=0,1,\ldots,p \Bigr\}.
\end{equation}
Let
\[
 \xi=(\xi_0,\ldots,\xi_p), \quad \xi_i = Z_k\Bigl(t+\sum_{l=1}^i D_l e_l\Bigr)
\]
be the values of the random field $Z_k$ at the $p+1$ vertices of the simplex
 (\ref{simplex}).
This is a Gaussian random vector with a mean 0 and a covariance matrix
\begin{equation}
\label{sigma}
\Sigma
 =
\begin{pmatrix}
1 & \tau_1 & \tau_1\tau_2 & \cdots & \tau_1\tau_2\tau_3\cdots\tau_p \\
  & 1      & \tau_2       & \cdots & \tau_2\tau_3\cdots\tau_p \\
  &        & 1            & \cdots & \tau_3\cdots\tau_p \\
  &        &              & \ddots & \vdots \\
  &        &              &        & 1
\end{pmatrix}_{(p+1)\times(p+1)},
\end{equation}
where $\tau_i=\Cov(Z_k(t),Z_k(t+D_i e_i))=R_{ki}(D_i)$.
 (Although $\xi$ and $\tau_i$ depend on $k$, we omit the index $k$ for simplicity.)
We can define the random field $\widetilde Z_k$ by interpolating
the random vector $\xi$ into the simplex (\ref{simplex}).
To be precise, by the affine bijection map
from the canonical $p$-dimensional simplex
\[
 \Delta^p
 = \conv\{ 0, e_1,\ldots, e_p \}
 = \Bigl\{ s\in\R^p \mid 0\le s_i,\,\sum_i s_i\le 1 \Bigr\} 
\]
to the simplex (\ref{simplex}), we can introduce
a parameter (local coordinates) $s=(s_i)$ into (\ref{simplex}),
and define a Gaussian random field by
\[
 \widetilde Z_k(s) = \frac{(1-\sum_i s) \xi_0 + \sum_i s_i \xi_i}{\sigma(s)},
\]
where
\[
 \sigma(s) = \sqrt{\varphi(s)^\top\Sigma\varphi(s)}, \qquad \varphi(s) = \Bigl( 1-\sum_i s_i,s_1,\ldots,s_p \Bigr)^\top
\]
is the normalizing constant so that the variance of $\widetilde Z_k(s)$ is 1.

\subsubsection{Volume of the index set of the chi-square random fields}

The volume of the index set $\widetilde T\times\S^{m-1}$ can be obtained by
summing up the volumes of the index sets $\Delta^p\times\S^{m-1}$
for the Gaussian random fields
\[
 \widetilde X(s,u) = \sum_{k=1}^m u_k \widetilde Z_k(s),
 \quad (s,u)\in \Delta^p\times\S^{m-1}.
\]

Let $u=u(\theta_a)$ be a local coordinate of $\S^{m-1}$.
Partial derivatives with respect to $s_i$ and $\theta_a$ are denoted by
$\partial_i$ and $\partial_a$, respectively.
The covariance matrix of
\[
 \partial_i \widetilde X(s,u)
 = \sum_{k=1}^m u_k \partial_i \widetilde Z_k(s), \qquad
\partial_a \widetilde X(s,u) = \sum_{k=1}^m \partial_a u_k \widetilde Z_k(s)
\]
is
\[
 \begin{pmatrix}
 \sum_{k=1}^m u_k^2 g_{k,ij}(s) & 0 \\ 0 & \bar g_{ab}(u)
 \end{pmatrix},
\]
where
\[
 g_{k,ij}(s) = E[\partial_i Z_k(s) \partial_j Z_k(s)], \quad \bar g_{ab}(u) = \sum_{k=1}^m \partial_a u_k \partial_b u_k.
\]
Hence, the volume of the index manifold $\Delta^p\times\S^{m-1}$ is
\[
 \Vol(\Delta^p\times\S^{m-1}) = \int_{\Delta^p\times\S^{m-1}} C(s,u),
\]
where
\[
 C(s,u) = \det\Bigl(\sum_{k=1}^m u_k^2 g_{k,ij}(s)\Bigr)^{1/2}
 \prod_i ds_i \, du, \quad
 du = \det\bigl(\bar g_{ab}(u)\bigr)^{1/2}\prod_a d\theta_a
\]
is the volume element.

We consider the case where $D_i\sim 0$, or equivalently $\tau_i\sim 1$,
in $\Sigma$ (\ref{sigma}).
Let $J$ be the $(p+1)\times(p+1)$ matrix whose elements are all 1.
Then,
\[
 \Sigma = J - \Sigma_1 + O(\max |1-\tau_i|^2),
\]
where $\Sigma_1$ is a symmetric matrix such that
\[
 (\Sigma_1)_{ii}=0, \quad (\Sigma_1)_{ij} = \sum_{l=i}^{j-1} (1-\tau_l) \quad (i<j).
\]

By using the covariance function
\[
 \widetilde r_k(s,s') =
 \Cov\bigl(\widetilde Z_k(s),\widetilde Z_k(s')\bigr) =
 \frac{\varphi(s)^\top\Sigma\varphi(s')}
      {\sqrt{\varphi(s)^\top\Sigma\varphi(s)\cdot
             \varphi(s')^\top\Sigma\varphi(s')}},
\]
the metric of the index set $\Delta^p$ is induced by
\[
 g_k(s) = (g_{k,ij}(s))_{1\le i,j\le d}, \quad
 g_{k,ij}(s) = \frac{\partial^2 \widetilde r_k(s,s')}
                    {\partial s_i \partial s'_j}\Big|_{s'=s}.
\]

Simple calculations yield
\[
g_{k,ij}
= 
 \frac{\varphi_i^\top\Sigma\varphi_j}{\varphi^\top\Sigma\varphi}
-
 \frac{(\varphi_i^\top\Sigma\varphi)(\varphi_j^\top\Sigma\varphi)}
      {(\varphi^\top\Sigma\varphi)^2},
\]
\[
 \varphi_i = \frac{\partial\varphi(s)}{\partial s_i} = (-1,\underbrace{0,\ldots,0}_{i-1},1,\underbrace{0,\ldots,0}_{p-i})^\top.
\]
Abbreviating $O(\max |1-\tau_i|)$ as $O$ yields
\begin{align*}
\varphi^\top\Sigma\varphi =& \varphi^\top J \varphi + O = 1 + O, \quad
\varphi^\top\Sigma\varphi_j = \varphi^\top J \varphi_j + O = O, \\
\varphi_i^\top\Sigma\varphi_j
 =& \varphi_i^\top J \varphi_j - \varphi_i^\top \Sigma_1 \varphi_j +O^2
 = - \varphi_i^\top \Sigma_1 \varphi_j + O^2 \\
 =& - (\Sigma_1)_{11} + (\Sigma_1)_{i+1,1}
    + (\Sigma_1)_{1,j+1} - (\Sigma_1)_{i+1,j+1} + O^2 \\
 =&
 \begin{cases}
 \sum_{l=1}^{i} (1-\tau_l) + \sum_{l=1}^{j} (1-\tau_l)
   - \sum_{l=i+1}^{j} (1-\tau_l) + O^2 & (i<j), \\
  2 \sum_{l=1}^{i} (1-\tau_l) + O^2 & (i=j) \end{cases} \\
 =& 2 \sum_{l=1}^{i} (1-\tau_l) + O^2 \quad (i\le j),
\end{align*}
and
\[
 g_{k,ij} = \Bigl\{ 2 \sum_{l=1}^{\min(i,j)} (1-\tau_l) \Bigr\}
 (1 + O(\max |1-\tau_i|)).
\]
By substituting
$\tau_i = 1 - \rho_{ki} D_i + o(D_i)$, we obtain
\[
 g_{k,ij} = \Bigl( 2 \sum_{l=1}^{\min(i,j)} \rho_{kl} D_l \Bigr) (1 + o(1))
 \quad (\max D_i \to 0).
\]
Some simple calculations yield
\begin{align*}
\det\Bigl(\sum_{k=1}^m u_k^2 g_{k,ij}(s)\Bigr)^{1/2}
&= \det\Bigl(2 \sum_{l=1}^{\min(i,j)}
 \Bigl( \sum_{k=1}^m u_k^2 \rho_{kl} \Bigr) D_l \Bigr)^{1/2} (1+o(1)) \\
&= 2^{p/2} \prod_{i=1}^p D_i^{1/2} \prod_{i=1}^p \bar\rho_i(u)^{1/2} (1+o(1)),
\end{align*}
where $\bar\rho_i(u)$ is defined in (\ref{rhobar}).
Combined with
\[
 \int_{\Delta^p} \prod_i ds_i
 = \int_{ 0\le s_i,\,\sum s_i\le 1 } \prod_i ds_i = \frac{1}{p!},
\]
we obtain the volume of the index set $\Delta^p\times\S^{m-1}$ as
\begin{equation}
\label{vol0}
 \frac{2^{p/2} C}{p!} \prod_{i=1}^p D_i^{1/2} (1+o(1)), \qquad
 C = \int_{\S^{m-1}}
 \prod_{i=1}^p \bar\rho_i(u)^{1/2} \, du.
\end{equation}

By letting $D_i:=D_{i j_i}$, and summing up (\ref{vol0})
with respect to $j_i=1,\ldots,n_i$ ($i=1,\ldots,p$),
we can show that the volume of $\widetilde T\times\S^{m-1}$ is
\[
 \Vol(\widetilde T\times\S^{m-1}) =
 2^{p/2} C \prod_{i=1}^p \Bigl(\sum_{j=1}^{n_i} D_{ij}^{1/2}\Bigr) (1+o(1)).
\]
By substituting this into (\ref{tube0}),
we obtain the tube formula (\ref{tube}) for the probability
$P\bigl(\max_{t\in \widetilde T} \widetilde Y(t) \ge b\bigr)$.

\begin{acknowledgements}
The authors thank David Siegmund and I-Ping Tu for helpful comments,
Matt Shenton for careful proofreading,
and Yuki Hasebe for help in preparing figures.
This work was supported by the Systems Genetics Project of the Research Organization of Information and Systems.
\end{acknowledgements}


\begin{thebibliography}{99}

\bibitem[Adler and Taylor(2007)]{Adler-Taylor07}
%
{\sc Adler, R.\,J.\ and Taylor, J.\,E.} (2007). 
{\it Random Fields and their Geometry\/}, 
Springer, New York. 

\bibitem[Agresti(2002)]{Agresti02}
%
{\sc Agresti, A.} (2002).
{\it Categorical Data Analysis\/}, 2nd ed,
Wiley-Interscience, Hoboken, New Jersey.

%
\bibitem[Bikard, et~al.(2009)]{Bikard-etal09}
%
{\sc Bikard, D., Patel, D., Le Mett\'{e}, C., Giorgi, V., Camilleri, C., Bennett, M.\,J., and Loudet, O.} (2009).
Divergent evolution of duplicate genes leads to genetic incompatibilities
 within {\it A.\,thaliana\/},
{\it Science\/}, {\bf 323}\ (5914), 623--626.

%
\bibitem[Churchill and Doerge(1994)]{Churchill-Doerge94}
%
{\sc Churchill, G.\,A.\ and Doerge, R.\,W.} (1994).
Empirical threshold values for quantitative trait mapping,
{\it Genetics\/}, {\bf 138} (3), 963--971.

%
\bibitem[Coyne and Orr(2004)]{Coyne-Orr04}
%
{\sc Coyne, J.\,A.\ and Orr, H.\,A.} (2004).
{\it Speciation\/}, 
Sinauer Associates, Sunderland.
%

%
\bibitem[Dobzhansky(1951)]{Dobzhansky51}
%
{\sc Dobzhansky, T.} (1951).
{\it Genetics and the Origin of Species\/}, 3rd ed., revised,
%
Columbia Univ.\ Press, New York.

%
\bibitem[Dupuis and Siegmund(1999)]{Dupuis-Siegmund99}
%
{\sc Dupuis, J.\ and Siegmund, D.} (1999).
Statistical methods for mapping quantitative trait loci from a dense set
 of markers,
{\it Genetics\/}, {\bf 151}\ (1), 373--386.

%
\bibitem[Dupuis and Siegmund(2000)]{Dupuis-Siegmund00}
%
{\sc Dupuis, J.\ and Siegmund, D.} (2000).
Boundary crossing probabilities in linkage analysis,
in {\it Game Theory, Optimal Stopping, Probability and Statistics:
 Papers in honor of Thomas S.\ Ferguson\/},
F.\,T.\ Bruss and L.\ Le Cam, eds.,
IMS Lecture Notes --- Monograph Series, {\bf 35}, Beachwood: IMS, 141--152.

%
\bibitem[Harushima, et~al.(1998)]{Harushima-etal98}
%
{\sc Harushima, Y., Yano, M., Shomura, A., Sato, M., Shimano, T., Kuboki, Y., Yamamoto, T., Lin, S.\,Y., Antonio, B.\,A., Parco, A., Kajiya, H., Huang, N., Yamamoto, K., Nagamura, Y., Kurata, N., Khush, G.\,S., and Sasaki, T.} (1998).
A high-density rice genetic linkage map with 2275 markers using a single
 $\mathrm{F}_2$ population,
{\it Genetics\/}, {\bf 148}\ (1), 479--494. \\
Data are available at 
 {\tt http://rgp.dna.affrc.go.jp/pub/geneticmap98/}

\bibitem[Hirotsu(1997)]{Hirotsu97}
{\sc Hirotsu, C.} (1997).
Two-way change point model and its application,
{\it Austral.\ J.\ Statist.\/}, {\bf 39} (2), 205--218.

%
\bibitem[Kao, et al.(2010)]{Kao-etal10}
%
{\sc Kao, K.\,C., Schwartz, K., and Sherlock, G.} (2010).
A genome-wide analysis reveals no nuclear Dobzhansky-Muller pairs
of determinants of speciation between
{\it S.\,cerevisiae} and {\it S.\,paradoxus\/},
but suggests more complex incompatibilities,
{\it PLoS Genetics\/}, {\bf 6}, e1001038.

%
\bibitem[Kim and Siegmund(1989)]{Kim-Siegmund89}
%
{\sc Kim, H.-J.\ and Siegmund, D.} (1989).
The likelihood ratio test for a change-point in simple linear regression,
{\it Biometrika\/}, {\bf 76}\ (3), 409--423.

%
\bibitem[Kuriki and Takemura(2001)]{Kuriki-Takemura01}
%
{\sc Kuriki, S.\ and Takemura, A.} (2001).
Tail probabilities of the maxima of multilinear forms and their applications,
{\it Ann.\ Statist.\/}, {\bf 29}\ (2), 328--371.

%
\bibitem[Kuriki and Takemura(2009)]{Kuriki-Takemura09}
%
{\sc Kuriki, S.\ and Takemura, A.} (2009).
Volume of tubes and distribution of the maxima of Gaussian random fields,
{\it Selected Papers on Probability and Statistics\/},
American Mathematical Society Translations Series 2,
Providence, Rhode Island: AMS, 25--48.

%
\bibitem[Mayr(1942)]{Mayr42}
%
{\sc Mayr, E.} (1942).
{\it Systematics and the Origin of Species
 from the Viewpoint of a Zoologist\/}, 
Columbia University Press, New York.

%
\bibitem[Mizuta, et~al.(2010)Mizuta, Harushima and Kurata]{Mizuta-etal10}
%
{\sc Mizuta, Y., Harushima, Y., and Kurata, N.} (2010).
Rice pollen hybrid incompatibility caused by reciprocal gene loss of duplicated genes,
{\it Proc.\ Natl.\ Acad.\ Sci.\ USA\/}, {\bf 107} (47), 20417--20422.

%
\bibitem[Ninomiya(2004)]{Ninomiya04}
%
{\sc Ninomiya, Y.} (2004).
Construction of conservative testing for change-point problems in
 two-dimensional random fields,
{\it J.\ Multivariate Anal.\/}, {\bf 89}\ (2), 219--242.

%
\bibitem[Piterbarg(1996)]{Piterbarg96}
%
{\sc Piterbarg, V.\,I.} (1996).
{\it Asymptotic Methods in the Theory of Gaussian Processes and Fields\/},
Translations of Mathematical Monographs, {\bf 148}, 
AMS, Providence, Rhode Island.

%
\bibitem[Reba\"i, et~al.(1994)]{Rebai-etal94}
%
{\sc Reba\"i, A., Goffinet, B., and Mangin, B.} (1994).
Approximate thresholds of interval mapping tests for QTL detection,
{\it Genetics\/}, {\bf 138}\ (1), 235--240.

%
\bibitem[Siegmund(1985)]{Siegmund85}
%
{\sc Siegmund, D.} (1985).
{\it Sequential Analysis: Tests and Confidence Intervals\/}, 
Springer, New York.

%
\bibitem[Siegmund(1988)]{Siegmund88}
%
{\sc Siegmund, D.} (1988).
Approximate tail probabilities for the maxima of some random fields,
{\it Ann.\ Probab.\/}, {\bf 16}\ (2), 487--501.

%
\bibitem[Siegmund(1992)]{Siegmund92}
%
{\sc Siegmund, D.\,O.} (1992).
Tail approximations for maxima of random fields, 
in {\it Probability Theory\/}, L.\,H.\,Y.\ Chen, K.\,P.\ Choi, K.\,Hu and 
J-H.\ Lou (eds.), Berlin: Walter de Gruyter, 147--158.
%

%
\bibitem[Siegmund(2004)]{Siegmund04}
%
{\sc Siegmund, D.} (2004).
Model selection in irregular problems:
 Applications to mapping quantitative trait loci,
{\it Biometrika\/}, {\bf 91}\ (4), 785--800.

%
\bibitem[Siegmund and Yakir(2007)]{Siegmund-Yakir07}
%
{\sc Siegmund, D.\ and Yakir, B.} (2007).
{\it The Statistics of Gene Mapping\/}, 
Springer, New York.

%
\bibitem[Sun(1993)]{Sun93}
%
{\sc Sun, J.} (1993).
Tail probabilities of the maxima of Gaussian random fields,
{\it Ann.\ Probab.\/}, {\bf 21} (1), 34--71.

%
\bibitem[Takemura and Kuriki(2002)]{Takemura-Kuriki02}
%
{\sc Takemura, A.\ and Kuriki, S.} (2002).
On the equivalence of the tube and Euler characteristic methods for the
 distribution of the maximum of Gaussian fields over piecewise smooth domains,
{\it Ann.\ Appl.\ Probab.\/}, {\bf 12}\ (2), 768--796.

%
\bibitem[Takemura and Kuriki(2003)]{Takemura-Kuriki03}
%
{\sc Takemura, A.\ and Kuriki, S.} (2003).
Tail probability via tube formula when critical radius is zero,
{\it Bernoulli\/}, {\bf 9}\ (3), 535--558.

%
\bibitem[Woodroofe(1982)]{Woodroofe82} 
%
{\sc Woodroofe, M.} (1982).
{\it Nonlinear Renewal Theory in Sequential Analysis\/},
CBMS-NSF Regional Conference Series in Applied Mathematics, {\bf 39}, 
SIAM, Philadelphia.

\end{thebibliography}
\end{document}